\documentclass[fleqn,usenatbib]{mnras}

\usepackage{newtxtext,newtxmath}

\usepackage[T1]{fontenc}

\DeclareRobustCommand{\VAN}[3]{#2}
\let\VANthebibliography\thebibliography
\def\thebibliography{\DeclareRobustCommand{\VAN}[3]{##3}\VANthebibliography}

\usepackage{graphicx}	
\usepackage{amsmath}	
\usepackage{fix-cm}

\title[Foreground component mapping with REACH]{Synchrotron and free-free mapping with simulated REACH observations between 50-170~MHz
}

\author[Daniel Robins et al.]{
Daniel Robins,$^{1,2}$\thanks{E-mail: dlr38@cam.ac.uk}
Dominic Anstey,$^{1,2}$
Harry Bevins,$^{1,2}$
Eloy de Lera Acedo$^{1,2}$
and Melis O. Irfan$^{3}$
\\
$^{1}$Astrophysics Group, Cavendish Laboratory, J. J. Thomson Avenue, Cambridge CB3 0HE, UK\\
$^{2}$Kavli Institute for Cosmology, Madingley Road, Cambridge CB3 0HA, UK\\
$^{3}$Institute of Astronomy, Madingley Road, Cambridge CB3 0HA, UK
}

\date{Accepted XXX. Received YYY; in original form ZZZ}

\pubyear{2026}

\begin{document}
\label{firstpage}
\pagerange{\pageref{firstpage}-\pageref{lastpage}}
\maketitle

\begin{abstract}
Global 21cm experiments aim to detect the hydrogen 21cm signal by separating it from foreground emission that can be orders of magnitude brighter than the signal.
REACH (the Radio Experiment for the Analysis of Cosmic Hydrogen) forward-models the sky by jointly fitting signal and foreground spectral parameters to an existing sky map.
The fitted parameters yield spectrally constrained, absolutely calibrated maps of the radio sky across the full 50-170~MHz observing band, among the lowest continuous frequencies yet mapped.
We assess REACH's ability to fit the 21cm signal and recover accurate foreground maps, using physically motivated foreground models of increasing complexity (starting from a pure synchrotron power law model, then introducing variable amplitudes, curvature, and a free-free component).
We evaluate these models against simulated REACH observations of correspondingly complex foregrounds, based on the Global Sky Model and the Python Sky Model.
To recover the 21cm signal, more complex datasets require correspondingly complex models, but this introduces degeneracies which limit accurate recovery of foreground parameters.
Fitting a foreground with independent synchrotron and free-free emission enables component-separated sky mapping, which has applications beyond radio cosmology; synchrotron is well-recovered across the sky, but free-free recovery is limited.
REACH is therefore capable of probing Galactic physics at uniquely low frequencies, alongside its primary goal of detecting the 21cm signal.
\end{abstract}





\section{Introduction}

The detection of the 21cm hydrogen signal from the Epoch of Reionization (EoR) and Cosmic Dawn is a key observational goal in modern cosmology \citep{Furlanetto2006,Morales2010,Pritchard2012,Furlanetto2016}.

The 21cm signal arises from the hyperfine spin-flip transition of neutral hydrogen, appearing in absorption or emission against the CMB depending on the relative values of the hydrogen spin temperature and the CMB temperature \citep{chapman2019cosmic}.
Its evolution is governed by astrophysical processes associated with the first luminous sources, including Lyman~$\alpha$ coupling, X-ray heating of the intergalactic medium (IGM), and reionization \citep{Furlanetto2006,Pritchard2012}.

Redshifted to within the $\sim$50-170 MHz (redshift $28\gtrsim z \gtrsim 7$) frequency band (assuming standard $\Lambda$CDM cosmology), the 21cm signal encodes the thermal and ionization history of the early Universe.
This includes information on the properties of the first stars and black holes - such as formation efficiency, timing, and initial mass function \citep[e.g.][]{Yajima2014,Mirocha2019,GesseyJones2022} - as well as the Lyman~$\alpha$ radiation field \citep{mittal2022implications} and heating of the IGM by X-ray binaries and cosmic rays \citep{Fialkov2014,GesseyJones2023}.
It also probes high-redshift radio backgrounds and cooling mechanisms \citep{Mebane2020}, and possible exotic physics such as dark matter-baryon interactions \citep{Barkana2018}.

Despite its scientific value, the 21cm signal has not yet been reliably detected; the claimed detection by EDGES \citep{Bowman2018} generated significant interest, but its interpretation remains uncertain \citep[see e.g.][]{Hills2018,Sims2020,Singh2022}.
A key obstacle in the reliable detection of the 21cm signal is the separation of the signal from Galactic and extragalactic foregrounds \citep{mittal2024impact}, which can be more than four orders of magnitude brighter than the 21cm signal \citep{Santos2005,Liu2020,Anstey2021}.
This makes accurate foreground characterization and removal essential for any 21cm experiment \citep{shaver1999can}.

When the foreground is characterised using forward-modelling, the fitted parameters yield spectrally constrained corrections and extensions to an existing sky map across the full observing band.
At the low frequencies relevant to 21cm experiments, there are few such maps in existence.
The 150\,MHz all-sky survey of \cite{Landecker1970} and earlier northern-sky maps at 178\,MHz \citep{Turtle1962} and 404\,MHz \citep{PaulinyToth1962} established early absolutely calibrated reference points for Galactic emission.
The landmark 408\,MHz survey of \cite{Haslam1982}, reprocessed in \cite{Remazeilles2015}, is the most widely-used radio sky map; its brightness temperature scale was tied to \cite{PaulinyToth1962}, with an overall scale uncertainty of $\sim$5-10\%.
The 22\,MHz survey of \cite{Roger1999} and near-complete 45\,MHz all-sky map of \cite{guzman2011all} extend coverage to lower frequencies, while more recently, digital aperture-synthesis arrays have produced multi-frequency maps at arcminute resolution, such as the LWA1 Low Frequency Sky Survey \citep{Dowell2017} at individual frequencies between 35-80\,MHz and the OVRO-LWA maps between 36-73\,MHz \citep{Eastwood2018}.

The Global Sky Model \citep[GSM;][]{deOliveiraCosta2008,Zheng2017} is a commonly-used model of the low-frequency radio sky, which is constructed by performing principal component analysis on a set of radio surveys and interpolating the resulting component weights to arbitrary frequencies.

The Python Sky Model \citep[PySM;][]{Thorne2017} is an alternative to the GSM widely used by the CMB community.
PySM simulates each Galactic emission component separately using physical scaling laws anchored to template maps derived from data including the Planck and WMAP missions, and other surveys \citep[including][]{Haslam1982}.
Unlike the GSM, PySM is physically motivated in its component separation, but both approaches involve frequency extrapolation.
PySM offers a different foreground basis against which to test 21cm signal recovery, although this requires extrapolating well beyond the GHz-frequency surveys on which its templates are calibrated.

Recent measurements have demonstrated that many low-frequency sky maps, and hence the GSM, require significant corrections, resulting in offsets in the GSM greater than 100K below 100\,MHz \citep{wilensky2025bayesian,McKay2026}.
A sky map based on continuous-frequency measurements, with reliable absolute calibration, would therefore be of significant astrophysical value.
Global 21cm experiments can provide this: the fitted foreground parameters yield spectrally constrained maps across the full observing band, and the experiment is absolutely calibrated in brightness temperature, with calibration uncertainties targeted at the millikelvin level required for 21cm signal detection.
This is in addition to their primary cosmological goal of 21cm signal detection.

At these low frequencies, the foreground is dominated by Galactic synchrotron emission from relativistic electrons spiralling in the Galactic magnetic field, with a subdominant contribution from Galactic free-free (thermal bremsstrahlung) emission, which is expected to contribute at the few per cent level \citep{Davies1998,Santos2005,Condon2016}.
Thermal dust emission, which dominates at frequencies above $\sim$10\,GHz, and anomalous microwave emission, which peaks around $\sim$30~GHz in certain regions of the sky, are entirely negligible at the frequencies corresponding to Cosmic Dawn and the EoR \citep{shaver1999can}.

The Radio Experiment for the Analysis of Cosmic Hydrogen \citep{DeLeraAcedo2022} is a global 21cm experiment operating from 50-170\,MHz, designed to measure the sky-averaged 21cm signal.
Like other global experiments, REACH faces the fundamental challenge of separating a spectrally smooth foreground (which can be modelled) from the spectral structure of the cosmological 21cm signal.
The choice of foreground model is critical: it must be flexible enough to capture the true foreground behaviour, while being sufficiently constrained to avoid absorbing features of the cosmological 21cm signal.

Several foreground modelling approaches have been proposed and implemented in 21cm experiments.
Experiments such as EDGES \citep{Bowman2018} and SARAS \citep{Singh2022saras} fit polynomials and other smooth functions directly to the data, to represent the foreground contribution to the integrated sky temperature.
In contrast, REACH uses a forward-modelling framework \citep{Anstey2021} in which an existing sky map is scaled by a spectral function, then convolved with the antenna beam.
This incorporates spatial structure and physical emission processes into the model, but it requires an assumed sky map as an anchor.

In this forward modelling framework, the simplest model is a power law in frequency \citep{Santos2005,Harker2012,Anstey2021}, motivated by the dominance of synchrotron emission.
Natural extensions include models with variable amplitudes \citep{pagano2024general}, spectral curvature terms, and physically motivated decompositions into synchrotron and free-free emission components.
However, such models have not been applied to global 21cm experiments under realistic observing conditions, and we believe this to be the first time that a global 21cm experiment has been used to decompose synchrotron and free-free emission, even using simulated foregrounds.

The REACH pipeline models the sky by dividing it into spatial regions, each assigned its own set of spectral parameters \citep{Anstey2021}.
Within each region, the foreground spectrum can be described by different models of increasing complexity, and this work considers four such models:
\begin{enumerate}
\item A standard power law model (synchrotron only)
\item A variable amplitude power law model (synchrotron only)
\item A curved power law model (synchrotron only)
\item A non-curved synchrotron + free-free component model.
\end{enumerate}

Curved synchrotron spectra can arise from superposition of source populations with different spectral indices, or from energy-dependent electron propagation \citep{Kogut2012,Platania2003,adam2016planck}.
At REACH frequencies, spectral curvature is expected to be finite but small: \cite{Kogut2012} measured a curvature parameter $c\approx-0.052$ at 310\,MHz, and the spectrum flattens toward lower frequencies \citep{Roger1999,guzman2011all}.

An alternative class of foreground model uses polynomials in frequency, without reference to physical emission processes or spatial components \citep[e.g.][]{pritchard2010constraining,Bowman2018,Singh2022saras,hibbard2023fitting}.
A stricter variant is maximally smooth functions - polynomials constrained to have no inflection points or zero crossings in high-order derivatives \citep{sathyanarayana2015detection,sathyanarayana2017modeling} - which have been applied to 21cm signal recovery in \cite{bevins2021maxsmooth}.
In this work, we focus on physically motivated models in which spectral parameters vary across sky regions, as this allows the fitted parameters to be interpreted in terms of physical emission processes, and used to construct component-separated sky maps.

In Section~\ref{sec:methods} we describe the four foreground models used in this work, the simulated REACH observations against which they are tested, and the Bayesian inference framework used to fit them.
Our foreground models are based on the work presented in \cite{Anstey2021}, in which the authors divide the sky into regions with similar spectral indices.

In Section~\ref{sec:results}, we address three objectives.
Firstly, \textit{21cm signal recovery} (Section \ref{sec:signal_extraction}) assesses which model best extracts the 21cm signal from simulated observations generated with different foreground models.
Secondly, \textit{foreground parameter recovery} (Section \ref{sec:param_recovery}) assesses how accurately each model recovers true foreground parameters.
Accurate foreground parameter recovery is needed to construct accurate foreground sky maps.
Thirdly, \textit{component-separated mapping} (Section \ref{sec:component_maps}) is unique to the synchrotron + free-free model, in which synchrotron and free-free emission are fitted separately.

We demonstrate that physically motivated component models improve interpretability, but require tailored spatial region splitting to remain identifiable: the optimal region partition for each emission component depends on that component's spatial distribution.

We discuss the implications for 21cm experiments in Section~\ref{sec:discussion}, and conclude in Section~\ref{sec:conclusions}.

\section{Methodology}
\label{sec:methods}

\subsection{Foreground Models}
\label{sec:models}

To recover the 21cm signal, a foreground model must capture all foreground spectral structure but avoid absorbing 21cm signal features.
We compare four foreground models of increasing complexity, based on the framework of \cite{Anstey2021}, in which the sky is divided into $N_{\rm reg}$ regions by percentile splits \citep{pagano2024general} of the spectral-index map.
This ensures each region contains a similar sky area while grouping pixels with similar spectral properties.
We use $N_{\rm reg} = 10$ in our fiducial runs, comparing alternative values in Section~\ref{sec:complexity}.
Within each region $i$, the sky brightness temperature $T_{\rm sky,\mathit{i}}(\Omega,\nu)$ is described as a function of position $\Omega$ and frequency $\nu$, anchored to a reference sky map whose amplitude and spectral behaviour are allowed to vary between regions to fit the data.
We take this reference map to be an instance of the 2008 GSM \citep{deOliveiraCosta2008} at 230\,MHz, a frequency at which the map is free of 21cm signal; it does however include the CMB contribution $T_{\rm CMB} \approx 2.725$\,K.

\begin{figure*}
\includegraphics[width=\textwidth]{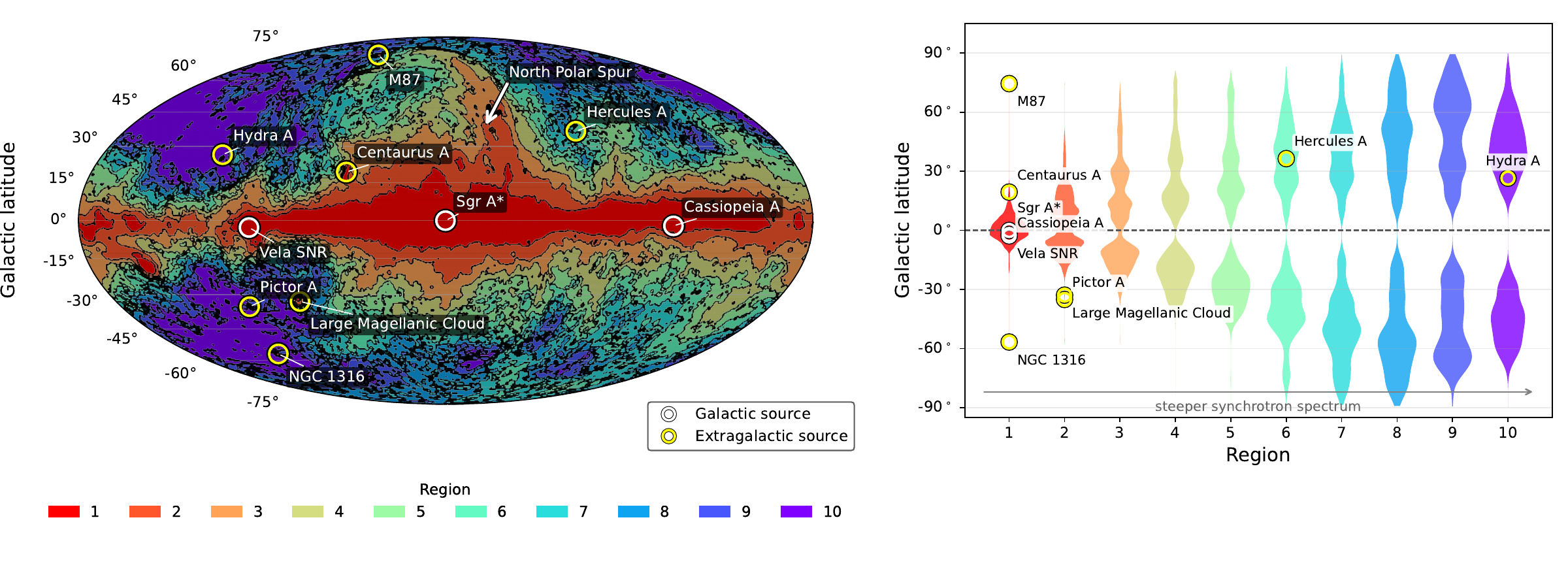}
\caption{Region definition and Galactic latitude coverage.
Left: sky map showing the 10 percentile-split regions.
Right: distribution of Galactic latitudes covered by each region, showing how the percentile-based division naturally separates extreme-latitude regions from regions near the Galactic plane.
In both panels, notable bright radio sources are labelled, including from Galactic and extragalactic sources \citep{green2025updated,helou1991nasa,Wenger2000}.}
\label{fig:region_latitude_explainer}
\end{figure*}

Figure~\ref{fig:region_latitude_explainer} shows how the 10 percentile-split regions divide the sky.
The right panel shows the distribution of Galactic latitudes covered by each region.
Various bright radio sources are identified, including supernova remnants \citep{green2025updated}, radio galaxies \citep{helou1991nasa}, and other prominent sources \citep{Wenger2000}.

While the percentile-based division broadly produces concentric regions, boundaries are influenced by local spectral-index structure: compact sources can shift nearby pixels into a different percentile bin, producing localised island regions isolated from the rest of their region.
To illustrate this, consider Region 1 in the right-hand plot in Figure~\ref{fig:region_latitude_explainer}: M87 and NGC~1316 are classified as Region 1, even though they occupy more extreme Galactic latitudes than the rest of this region.
This spectral-index-based grouping is advantageous for foreground modelling, as it groups pixels with similar spectral properties regardless of their Galactic latitude.

Within each region, the foreground spectrum is described by one of four models of increasing complexity, as follows.

\subsubsection{Power law model}

The standard power law model assumes a simple frequency ($\nu$) dependence at each point in the sky ($\Omega$):
\begin{equation}
T_{\rm sky,\mathit{i}}(\Omega, \nu) = [T_{230}(\Omega) - T_{\rm CMB}]\left(\frac{\nu}{230~{\rm MHz}}\right)^{-\beta_i} + T_{\rm CMB},
\label{eq:power}
\end{equation}
where $T_{230}(\Omega)$ is the sky brightness temperature at 230\,MHz, $T_{\rm CMB} \approx 2.725$\,K, and $\beta_i$ is the spectral index for region $i$.
Synchrotron emission, which dominates the low-frequency radio sky, follows an approximate power law, motivating this as the baseline model against which others are compared.
This model fits one parameter (spectral index $\beta$) per region, and is the approach taken by \cite{Anstey2021}.

\subsubsection{Variable amplitude power law model}

The variable amplitude power law extends the standard power law by introducing a region-dependent amplitude:
\begin{equation}
T_{\rm sky,\mathit{i}}(\Omega, \nu) = A_i[T_{230}(\Omega) - T_{\rm CMB}]\left(\frac{\nu}{230~{\rm MHz}}\right)^{-\beta_i} + T_{\rm CMB},
\label{eq:poweramp}
\end{equation}
where $A_i$ is a dimensionless amplitude parameter for region $i$ that scales the foreground contribution, correcting for errors in the absolute calibration of the 230\,MHz anchor map $T_{230}$ \citep[as in][]{pagano2024general}.
This gives $2N_{\rm reg}$ parameters in total, while retaining spectral smoothness.

\subsubsection{Curved power law}

The curved power law extends the variable amplitude power law by adding another spectral term that accounts for curvature:
\begin{equation}
\begin{aligned}
T_{\rm sky,\mathit{i}}(\Omega, \nu) = \; & A_i[T_{230}(\Omega) - T_{\rm CMB}] \\
& \times\left(\frac{\nu}{230~{\rm MHz}}\right)^{-\beta_i + c_i\ln(\nu/230~{\rm MHz})} + T_{\rm CMB},
\end{aligned}
\label{eq:curvature}
\end{equation}
where $A_i$ is the amplitude, $\beta_i$ the spectral index, and $c_i$ the curvature parameter for region $i$, giving $3N_{\rm reg}$ parameters total.
$c_i$ accounts for superpositions of radio sources with different spectral indices.

\subsubsection{Sync + ff model}

The sync + ff model explicitly decomposes the foreground into its two dominant physical components:
\begin{equation}
\begin{split}
T_{\rm sky,\mathit{i}}(\Omega, \nu) = [T_{230}(\Omega) - T_{\rm CMB}] &\left[A_{\rm sync,\mathit{i}}\left(\frac{\nu}{230~{\rm MHz}}\right)^{-\beta_{\rm sync,\mathit{i}}} \right. \\
&\left. + A_{\rm ff,\mathit{i}}\left(\frac{\nu}{230~{\rm MHz}}\right)^{-2.1}\right] + T_{\rm CMB},
\end{split}
\label{eq:syncff}
\end{equation}
where $A_{\rm sync,\mathit{i}}$ and $A_{\rm ff,\mathit{i}}$ weight the synchrotron and free-free contributions respectively, and $\beta_{\rm sync,\mathit{i}}$ is the uncurved synchrotron spectral index.
The free-free spectral index is fixed at its well-established value of $\beta_{\rm ff} = 2.1$, consistent with optically thin thermal bremsstrahlung \citep[e.g.][]{Rybicki1979,Dickinson2003,Draine2011,Xu2013,Condon2016}, giving $3N_{\rm reg}$ parameters in total.

This physical decomposition enables direct construction of separate synchrotron and free-free sky maps.
Throughout this work, the same region definitions are used for both components, but Section~\ref{sec:splitting} explores the alternative of defining synchrotron and free-free regions independently.

\subsubsection{Summary of models}

Table~\ref{tab:model_summary} summarises the key properties of the foreground models described above.

\begin{table*}
\centering
\caption{Summary of the four foreground models tested in this work.
$p_{\rm model}$ is the number of free parameters per region.}
\begin{tabular}{llll}
\hline
Model & $p_{\rm model}$ & Free parameters & Physical motivation \\
\hline
Power law & 1 & $\beta$ & Synchrotron emission, which dominates at low frequencies, obeys a power law.\\
Variable amplitude power law& 2 & $A$, $\beta$ & Brightness normalisation may vary spatially, due to imperfectly calibrated sky surveys.\\
Curvature& 3 & $A$, $\beta$, $c$ & Spectral curvature (e.g. from superposition of sources) may be non-zero.\\
Sync + ff& 3 & $A_{\rm sync}$, $\beta_{\rm sync}$, $A_{\rm ff}$ & Each emission component can be mapped separately.\\
\hline
\end{tabular}
\label{tab:model_summary}
\end{table*}

\subsection{Simulation Pipeline}
\label{sec:simulation}

We simulate REACH observations by convolving input sky maps with the REACH beam pattern and 
adding radiometric noise; full details are provided in \cite{Anstey2021}.
We simulate nine input sky maps with different foreground parametrisations (e.g.\ varying spectral curvature and free-free distribution), in order to test each foreground model against skies of corresponding complexity.
Eight of the nine maps are constructed using the spectral model forms defined in 
Section~\ref{sec:models}; the ninth is a Python Sky Model (PySM) realisation \citep{Thorne2017}, which uses a different spectral basis to the other maps.
We use the latest version of PySM, \texttt{PySM3} \citep{zonca2021python,pan2025full}.

Absolute sky maps are computed at HEALPix $N_{\rm side}=512$ resolution \citep{gorski2005healpix}; these are then converted into an `observed' sky in Section~\ref{sec:instrument}.
A Gaussian global 21cm absorption signal,
\begin{equation}
T_{21}(\nu) = -A\exp\!\left[-\frac{(\nu - f_0)^2}{2\sigma^2}\right],
\label{eq:injected_signal}
\end{equation}
with $\boldsymbol{\theta}_{21} = (f_0, \sigma, A) = (85\,\mathrm{MHz},\, 15\,\mathrm{MHz},\, 0.155\,\mathrm{K})$, is injected into every dataset.
Each simulated observation therefore contains foreground $+$ noise $+$ signal.

For datasets incorporating a non-uniform free-free emission template, the thermal component is computed from the emission measure map of \citet{Hutschenreuter2024} and the electron temperature map from the Planck Commander free-free product \citep{adam2016planck}, using a Gaunt-factor optical-depth model. 

\begin{figure}
\includegraphics[width=\columnwidth]{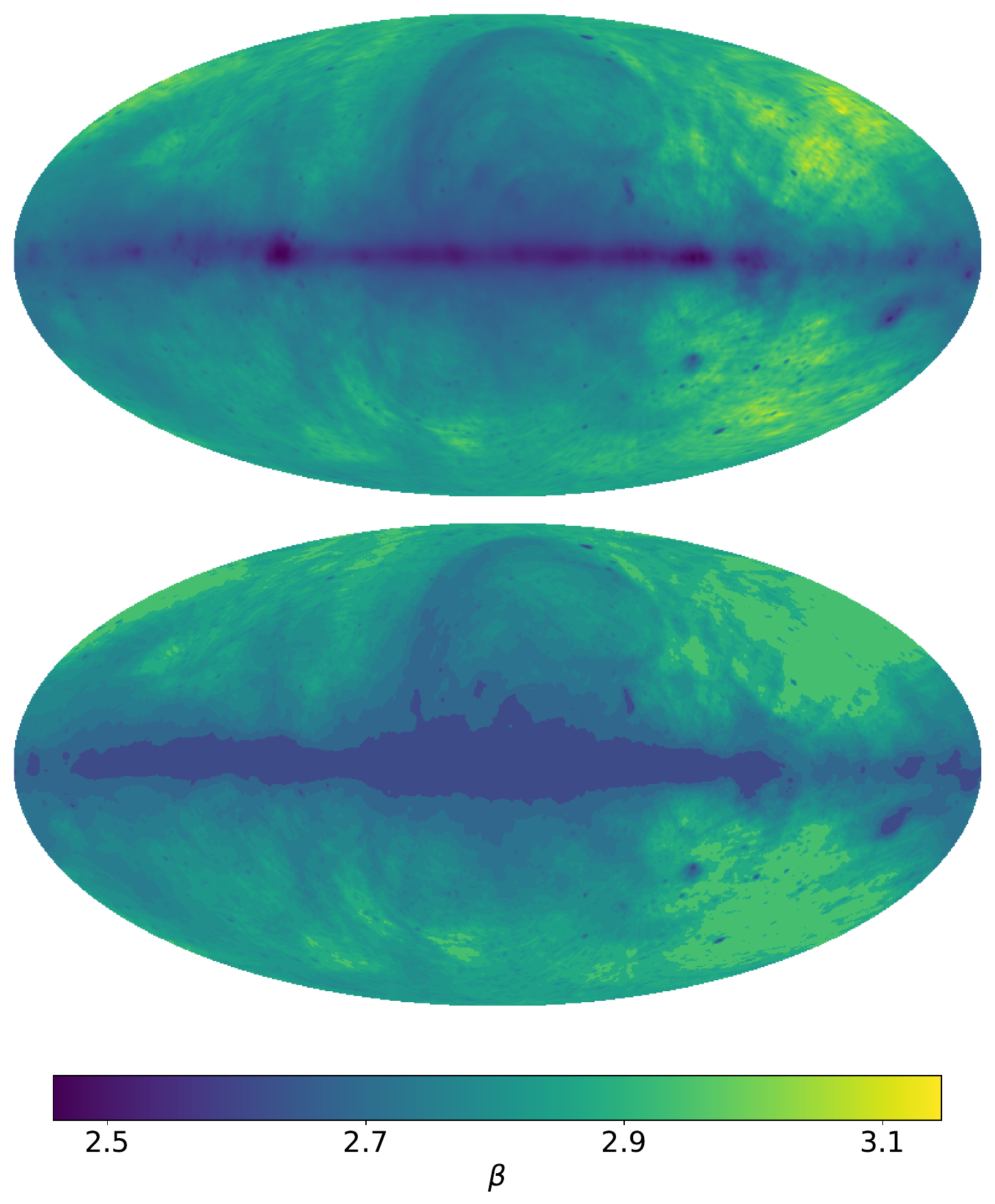}
\caption{Comparison of spectral index maps used for pixel-wise and region-wise data generation.
Top: continuous pixel-wise spectral indices from pixel-by-pixel scaling between an instance of the GSM \citep{deOliveiraCosta2008} at 230\,MHz and 408\,MHz, used for pixel-wise parameter data \citep{Anstey2021}.
Bottom: region-averaged spectral indices used for $N_{\rm reg} = 10$ region-wise parameter data, obtained by averaging the values of $\beta$ from the top panel within each region.}
\label{fig:spectral_index_comparison}
\end{figure}

We evaluate performance across two complementary data types: simulations using pixel-wise parameters (where the input parameters vary per pixel), and simulations using region-wise parameters (where the input parameters are uniform within each of the $N_{\rm reg} = 10$ regions).
Pixel-wise parameter data are more physically motivated and provide the most stringent test of model resilience.
Region-wise parameter data are designed for parameter recovery assessment, because there is a single true value of each input parameter.
Beam-smoothing of pixel-level variations may otherwise prevent meaningful comparison of fitted and true parameters.
As an example, Figure~\ref{fig:spectral_index_comparison} compares the pixel-wise and region-averaged spectral index maps obtained by scaling between the GSM at 408\,MHz and 230\,MHz \citep[as in][]{Anstey2021}.
It shows that the region-based approach simplifies parameter recovery assessment by eliminating beam-induced correlations between pixels, but also moves the most extreme values closer to the mean.

\begin{figure*}
\includegraphics[width=\textwidth]{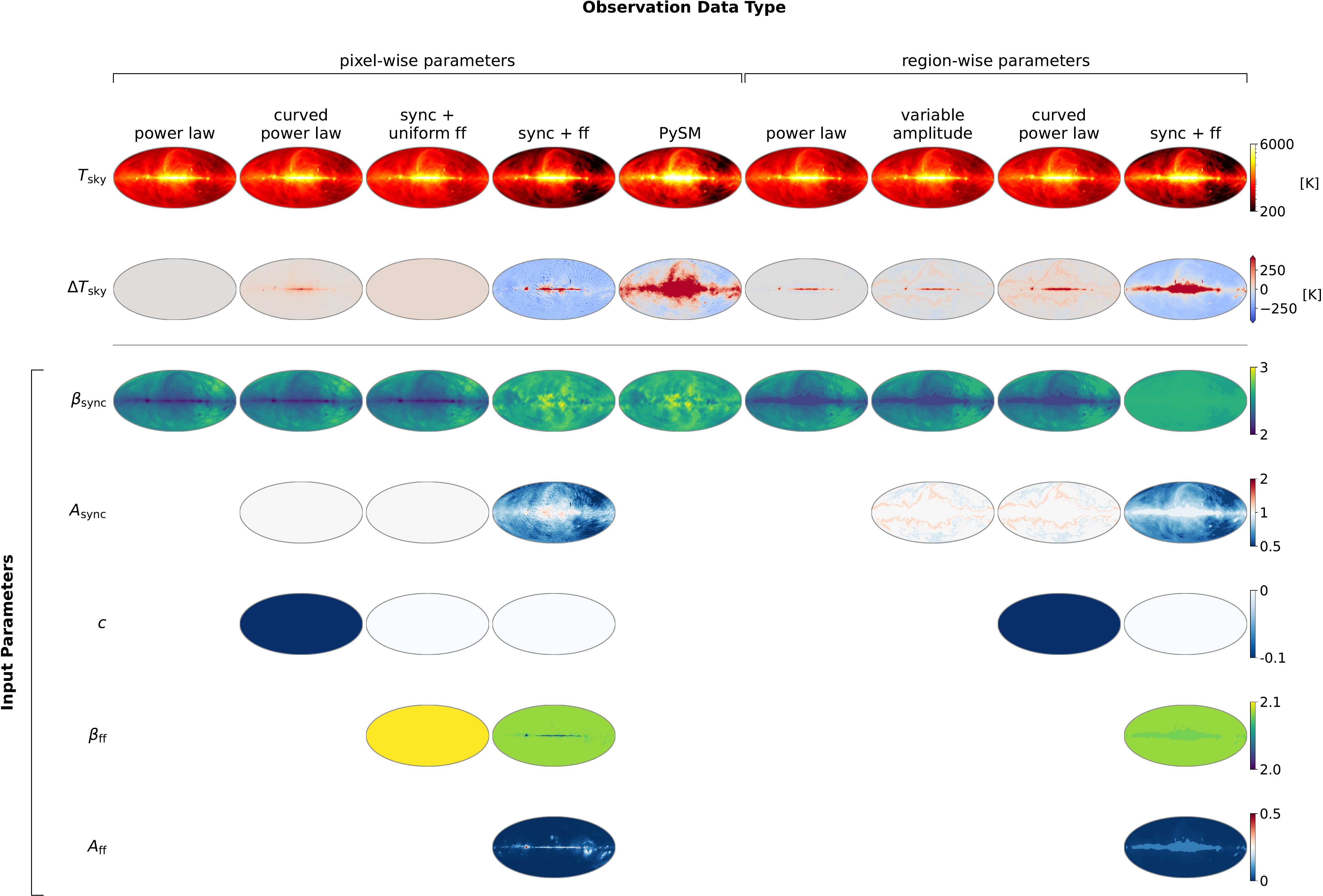}
\caption{Input parameter maps at 125\,MHz for all nine simulated datasets, including five pixel-wise datasets and four region-wise datasets.
The top row shows $T_{\rm sky}$ and the second row shows $\Delta T_{\rm sky}$ relative to the power law baseline.
Lower rows show the parameter maps used to generate each dataset: $\beta_{\rm sync}$, $A_{\rm sync}$, $\beta_{\rm ff}$, $A_{\rm ff}$, and $c$.
For models with fixed parameters, constant-value maps are shown; blank cells denote parameters not used in that observation data type.}
\label{fig:master_region_maps}
\end{figure*}

Figure~\ref{fig:master_region_maps} compares the input parameter maps for all nine foreground realisations at 125\,MHz, the details of which are discussed in Sections~\ref{sec:pixelwise} and \ref{sec:regionwise}.
The top two rows show the total sky temperature $T_{\rm sky}$ and the differential $\Delta T_{\rm sky}$ relative to the power-law baseline.
The lower rows show the parameter maps (where applicable) that were used to generate each dataset.
The \texttt{PySM3} dataset uses the \texttt{s1} and \texttt{f1} preset models, whose spectral parametrisation differs from the GSM-based models used for the other datasets, and so has no direct counterparts to these parameters.

The $T_{\rm sky}$ maps are taken as the true sky, against which we compare with the fitted sky map; these $T_{\rm sky}$ maps are \textit{not} inputted into the REACH data analysis pipeline.
More details on the construction of each observation data type are provided below.

\subsubsection{Pixel-wise parameter datasets}
\label{sec:pixelwise}
These five datasets use spatially varying parameters at the HEALPix pixel level $p$, providing the most physically realistic foreground realisations.

\begin{enumerate}
\item Power law: Standard pixel-wise power law using $T_{230}(p)$ and the spectral index map $\beta(p)$ obtained from a pixel-by-pixel scaling between an instance of the GSM at 230\,MHz and 408\,MHz (top panel of Figure \ref{fig:spectral_index_comparison}):
\begin{equation}
T_{\rm sky}(\nu,p) = \left[T_{230}(p) - T_{\rm CMB}\right]\left(\frac{\nu}{230\,{\rm MHz}}\right)^{-\beta(p)} + T_{\rm CMB}.
\label{eq:powerlaw}
\end{equation}

\item Curvature: Similar to the power law data, but with fixed spectral curvature $c = -0.1$ (motivated by the measurements in \cite{irfan2022measurements}):
\begin{multline}
T_{\rm sky}(\nu,p) = \left[T_{230}(p) - T_{\rm CMB}\right]
\left(\frac{\nu}{230\,{\rm MHz}}\right)^{-\beta(p)\,-0.1\ln(\nu/230\,{\rm MHz})} \\
+ T_{\rm CMB}.
\end{multline}

\item Synchrotron + uniform ff: Adds a spatially uniform free-free component on top of the power law synchrotron term given in equation~\ref{eq:powerlaw}, with a spectral index $\beta_{\rm ff} = 2.1$, fixed by a free-free brightness temperature of $T_{\rm ff,0} = 10$\,K at 230\,MHz:
\begin{multline}
T_{\rm sky}(\nu,p) = \left[T_{230}(p) - T_{\rm CMB}\right]
  \left(\frac{\nu}{230\,{\rm MHz}}\right)^{-\beta_{\rm sync}(p)} \\
  {} + T_{\rm ff,0}\left(\frac{\nu}{230\,{\rm MHz}}\right)^{-\beta_{\rm ff}} + T_{\rm CMB}.
\end{multline}

Note that $T_{\rm ff,0}$ and $A_{\rm ff}$ are not equivalent: $T_{\rm ff,0}$ is a single global parameter that sets the free-free brightness temperature across the whole sky, whereas $A_{\rm ff}(p)$ is a dimensionless pixel-wise scaling that modulates the free-free contribution relative to the local foreground amplitude at each pixel.
This dataset assumes free-free emission is present everywhere.
This is not physically realistic (thermal free-free emission is actually concentrated along the Galactic plane and in \ion{H}{ii} regions), but this data tests the robustness of each model when a spectrally distinct but spatially indiscriminate free-free component is added to the foreground, in a manner which is more extreme than reality.

\item Synchrotron + non-uniform ff: Physically motivated pixel-wise synchrotron + free-free decomposition, using observational maps to assign spatially varying amplitudes and spectral indices to each component.

The synchrotron and free-free amplitudes are defined as the fractional contribution of each component to the total foreground brightness at 230\,MHz,
\begin{equation}
A_{\rm sync}(p)=\frac{T_{\rm sync}(230\,{\rm MHz},p)}{T^{\rm total}_{230}(p)- T_{\rm CMB}},
\qquad
A_{\rm ff}(p)=\frac{T_{\rm ff}(230\,{\rm MHz},p)}{T^{\rm total}_{230}(p)- T_{\rm CMB}}.
\end{equation}

The synchrotron spectral index is obtained from the convolutional neural network (CNN)-simulated map $\beta_{\rm sync,CNN}(p)$ from \cite{Irfan2023}, which is derived from multi-frequency synchrotron separation analysis.
This has finer spatial structure than the GSM-derived spectral index used in the power law, curvature, and synchrotron + uniform ff datasets, with sub-degree variation modelled by CNNs trained on higher-resolution data.

The desourced, destriped Haslam 408\,MHz map \citep{Remazeilles2014} is then scaled by $\beta_{\rm sync,CNN}$ and $A_{\rm sync}$:
\begin{equation}
T_{\rm sync}(\nu,p)=
A_{\rm sync}(p)T_{408}(p)\,\left(\frac{\nu}{408\,{\rm MHz}}\right)^{-\beta_{\rm sync,CNN}(p)}.
\end{equation}

The free-free brightness temperature at three reference frequencies is computed from the emission measure map \citep{Hutschenreuter2024} and the Planck Commander electron-temperature map \citep{adam2016planck} via
\begin{equation}
T_{\rm ff}(\nu,p)=T_e(p)\left[1-\exp\!\big(-\tau(\nu,p)\big)\right],
\end{equation}
where the free-free optical depth $\tau$ is
\begin{equation}
\tau(\nu,p)=0.05468\,T_e(p)^{-1.5}
\left(\frac{\nu}{1000\,{\rm MHz}}\right)^{-2}
{\rm EM}(p)\,g_{\rm ff}(\nu,p),
\end{equation}
and the Gaunt factor $g_{\rm ff}$ is
\begin{equation}
g_{\rm ff}(\nu,p)=
\ln\!\left[
\exp\!\left(
5.960-\frac{\sqrt{3}}{\pi}\ln\!\left(\frac{\nu}{1000\,{\rm MHz}}
\left(\frac{T_e(p)}{10^4\,{\rm K}}\right)^{-1.5}\right)\right)+e
\right].
\end{equation}
$\beta_{\rm ff}(p)$, given by
\begin{equation*}
\beta_{\rm ff}(p) = \frac{\ln\bigl[T_{\rm ff}(260\,{\rm MHz},p)\,/\,T_{\rm ff}(200\,{\rm MHz},p)\bigr]}{\ln(260/200)},
\end{equation*}
scales between these reference frequencies.

\item Python Sky Model, \texttt{PySM3}: A mostly independent foreground realisation generated at HEALPix $N_{\rm side}=512$ and $1^{\circ}$ resolution using the \texttt{PySM3} preset models \texttt{s1} (synchrotron) and \texttt{f1} (free-free).

The \texttt{s1} model uses the desourced, destriped Haslam 408\,MHz map \citep{Remazeilles2014} as the synchrotron emission template, with a spatially varying spectral index derived from a combination of the Haslam 408\,MHz and WMAP 23\,GHz data \citep{bennett2013nine}.
A power-law frequency scaling is applied, with zero spectral curvature \citep{miville2008separation}.

The \texttt{f1} model follows the analytic Commander free-free prescription of \citet{Draine2011}, producing a degree-scale map of free-free emission anchored to 30\,GHz and scaled in frequency with a spatially constant spectral index of $-2.14$.
Small scale anisotropies were also added, at greater resolution than the surveys on which the model was based \citep{pan2025full}, so the \texttt{PySM3} sky has finer spatial structure than any of the other data types.

The total foreground map is then $T_{\rm fg}(\nu,p) = T_{\rm sync}(\nu,p) + T_{\rm ff}(\nu,p)$.
Because this construction uses physically motivated component-specific spectral models and input templates that differ from the GSM-based parametrisation used for the other datasets, it constitutes a different foreground realisation against which to test the REACH pipeline.
\end{enumerate}

\subsubsection{Region-wise parameter datasets}
\label{sec:regionwise}

The form of the region-wise data exactly matches the form of the foreground models, with a single ground-truth value for each parameter in each region, against which the model fit can be validated.

Except for the variable amplitude power law data, each of these data correspond to a pixel-wise data type.
In these cases, region parameters are derived by averaging the corresponding pixel-wise quantities within each region.
In all cases, these data use 10 percentile-split regions.

\begin{enumerate}
\item Power law: Each region $i$ is assigned $\beta_i = \langle\beta\rangle_i$ (mean of the GSM $\beta$ map within region $i$, shown in the lower panel of Figure \ref{fig:spectral_index_comparison}).

\item Variable amplitude power law: Same $\beta_i$ construction as the power law data, but each region is modified by an arbitrary amplitude $A_i = [1, 1, 1, 1.1, 1, 1, 1, 0.9, 1, 1]$ for regions $i = 1,\ldots,10$.
The non-unity amplitudes in regions 4 and 8 represent step-function gain errors, testing each model's robustness to multiplicative calibration errors in the input maps \citep{pagano2024general}.

\item Curved power law: Similar to the variable amplitude power law data, but with fixed spectral curvature $c = -0.1$ applied uniformly across all regions.

\item Synchrotron + non-uniform ff: For each region $i$, the three sync + ff parameters are region averages of the pixel-wise synchrotron + non-uniform ff maps: $\beta_{{\rm sync},\mathit{i}} = \langle\beta_{\rm sync}\rangle_i$, $A_{{\rm sync},\mathit{i}} = \langle A_{\rm sync}\rangle_\mathit{i}$, and $A_{{\rm ff},\mathit{i}} = \langle A_{\rm ff}\rangle_i$.
\end{enumerate}

\subsubsection{Instrument simulation}
\label{sec:instrument}

The simulated sky $T_{\rm sky}(\Omega, \nu, \boldsymbol{\theta}_{\rm fg})$, with foreground parameters $\boldsymbol{\theta}_{\rm fg}$, from Sections \ref{sec:pixelwise} and \ref{sec:regionwise} are convolved with a model of the frequency-dependent REACH log-spiral antenna beam $D(\Omega, \nu)$ \citep{dyson,anstey2022informing}, then rotated through the local sidereal time (LST) track and integrated to produce the antenna temperature at each frequency channel across the REACH band in 1\,MHz steps.

Each simulation starts at 2019-01-01 00:00:00 UTC, and integrates over a 12-hour observation at 1-min cadence.
In line with \cite{Anstey2021}, Gaussian antenna-temperature noise with standard deviation
\begin{equation}
\sigma_{\rm noise} = \frac{0.025}{\sqrt{N_t}}\,\mathrm{K},
\label{eq:noise}
\end{equation}
where $N_t=721$ is the number of time samples in the 12-hour LST integration, giving $\sigma_{\rm noise} \simeq 9.3\times10^{-4}$\,K, is added to each frequency channel.

At each time sample $t_k$ ($k = 1,\ldots,N_t$), the instantaneous beam-weighted antenna temperature is
\begin{equation}
    T_{\rm fg}(\nu, t_k, \boldsymbol{\theta}_{\rm fg})
    = \frac{\int D(\Omega,\nu)\, T_{\rm sky}(\Omega, \nu, t_k,
    \boldsymbol{\theta}_{\rm fg})\, \mathrm{d}\Omega}
    {\int D(\Omega,\nu)\, \mathrm{d}\Omega}.
    \label{eq:beam_convolution_instantaneous}
\end{equation}
The final simulated data is then the time-averaged antenna temperature over the observation (accounting for the rotation of the sky into the antenna frame):
\begin{equation}
    \bar{T}_{\rm fg}(\nu, \boldsymbol{\theta}_{\rm fg})
    = \frac{1}{N_t}\sum_{k=1}^{N_t} T_{\rm fg}(\nu, t_k, \boldsymbol{\theta}_{\rm fg}).
    \label{eq:beam_convolution}
\end{equation}
The observed (simulated) data are then
\begin{equation}
    T_{\rm data}(\nu)
    =
    \bar{T}_{\rm fg}(\nu,\boldsymbol{\theta}_{\rm fg, true})
    + T_{21 \rm , true}(\nu) + n(\nu),
\label{eq:observed_data}
\end{equation}
where $T_{21 \rm , true}(\nu)$ is the injected 21cm signal  from equation~(\ref{eq:injected_signal}) and $n(\nu)\sim\mathcal{N}(0,\sigma_{\rm noise}^2)$ is the noise realisation from equation~(\ref{eq:noise}).

\begin{figure}
\includegraphics[width=\columnwidth]{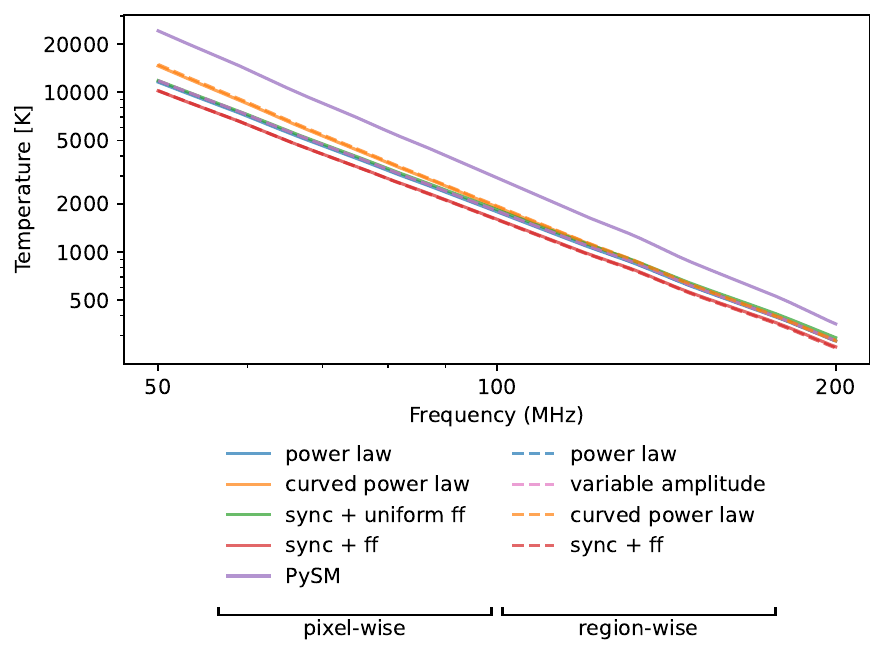}
\caption{Frequency spectra for all nine simulated observation datasets with 12-hour integration time.
All datasets include injected Gaussian noise and the 21cm signal of equation~(\ref{eq:injected_signal}).}
\label{fig:observation_data_comparison}
\end{figure}

Figure~\ref{fig:observation_data_comparison} shows these beam-weighted frequency spectra; it is these spectra that are input into the REACH data analysis pipeline, and not the maps generated in Sections~\ref{sec:pixelwise} and \ref{sec:regionwise}.

\subsection{Bayesian Inference Framework}
\label{sec:inference}

Full details of the Bayesian inference framework are found in \cite{Anstey2021,Anstey2023}, but we outline the key steps here.
The observed antenna temperature $T_{\rm data}(\nu)$ from equation~(\ref{eq:observed_data}) is fit with nested sampling \citep[via \texttt{PolyChord};][]{Handley2015a,Handley2015b}.

The likelihood compares $T_{\rm data}(\nu)$ to the fitted model via the difference
\begin{equation}
    \Delta_i \equiv T_{\rm data}(\nu_i)
        - T_{\rm fg}(\nu_i, \boldsymbol{\theta}_{\rm fg})
        - T_{21}(\nu_i, \boldsymbol{\theta}_{21}),
    \label{eq:data_model}
\end{equation}
with a Gaussian log-likelihood summed over frequency channels:
\begin{equation}
    \log\mathcal{L} = \sum_{i} \left[
        -\frac{1}{2}\log(2\pi\theta_{\sigma}^2)
        - \frac{\Delta_i^{2}}{2\theta_{\sigma}^2}
    \right],
    \label{eq:likelihood}
\end{equation}
where $\theta_{\sigma}$ is the noise standard deviation, treated as a free parameter of the fit.

The priors for the parameters in each model are provided in Appendix~\ref{app:priors}.

We run nested sampling until the evidence uncertainty is less than 0.001 (in log-evidence units).
For the four foreground models considered in this paper, the total number of sampled dimensions is
\begin{equation}
    N_{\rm dims} = p_{\rm model}\,N_{\rm reg} + N_{\rm signal} + N_{\rm like},
    \label{eq:ndims}
\end{equation}
where $p_{\rm model}$ is the number of free foreground parameters per region (1, 2, 3, and 3 for the power law, variable amplitude, curvature, and sync\,+\,ff models respectively), $N_{\rm signal} = 3$ for the three-parameter Gaussian signal profile from equation~(\ref{eq:injected_signal}), and $N_{\rm like} = 1$ for the Gaussian noise parameter.

\subsection{Comparison Metrics}
\label{sec:metrics}

We use several metrics to assess model performance.

The Bayesian evidence $\mathcal{Z}$ is the marginal likelihood of the data $\mathbf{d}$ under a given model with parameters $\boldsymbol{\theta}$, i.e.\ the probability of the data marginalised over the prior.
This allows the relative probabilities of different models to be compared.
$\mathcal{Z}$ is computed as
\begin{equation}
\mathcal{Z} = \int \mathcal{L}(\mathbf{d}|\boldsymbol{\theta}) \pi(\boldsymbol{\theta}) \, d\boldsymbol{\theta},
\end{equation}
where $\pi(\boldsymbol{\theta})$ is the prior.
The evidence naturally penalizes model complexity through the Occam factor inherent in Bayesian model comparison.

To assess 21cm signal recovery, we use the known true injected signal $T_{21,\rm true}(\nu)$ to calculate root mean square error (RMSE) between the weighted mean posterior signal $T_{21,\rm fit}(\nu)$ and the true signal at each frequency:
\begin{equation}
{\rm RMSE} = \sqrt{\frac{1}{N_{\rm data}}\sum_{i=1}^{N_{\rm data}}[T_{21,\rm fit}(\nu_i) - T_{21,\rm true}(\nu_i)]^2}
\end{equation}

To assess the recovery of foreground parameters in each region, only the region-wise data provide a fair assessment, as these have a single true parameter in each region.
So, we compare fitted parameter values $\hat{\theta}_i$ to true values $\theta_{\rm true, \mathit{i}}$ in each region $i$, weighted by the posterior standard deviation $\sigma_i$ of the fitted parameter in that region:

\begin{equation}
\Delta \theta_i = \frac{\hat{\theta}_i - \theta_{\rm true, \mathit{i}}}{\sigma_i}.
\end{equation}

We assess whether $\theta_{{\rm true, }\mathit{i}}$ falls within the fitted 1$\sigma$ credible interval for each parameter in each region.
This is particularly important for the sync + ff model, where accurate recovery of $A_{\rm sync}$, $A_{\rm ff}$, and $\beta_{\rm sync}$ is required for foreground mapping.

\subsection{Impact of Model Complexity}
\label{sec:complexity}

\begin{figure*}
\includegraphics[width=\textwidth]{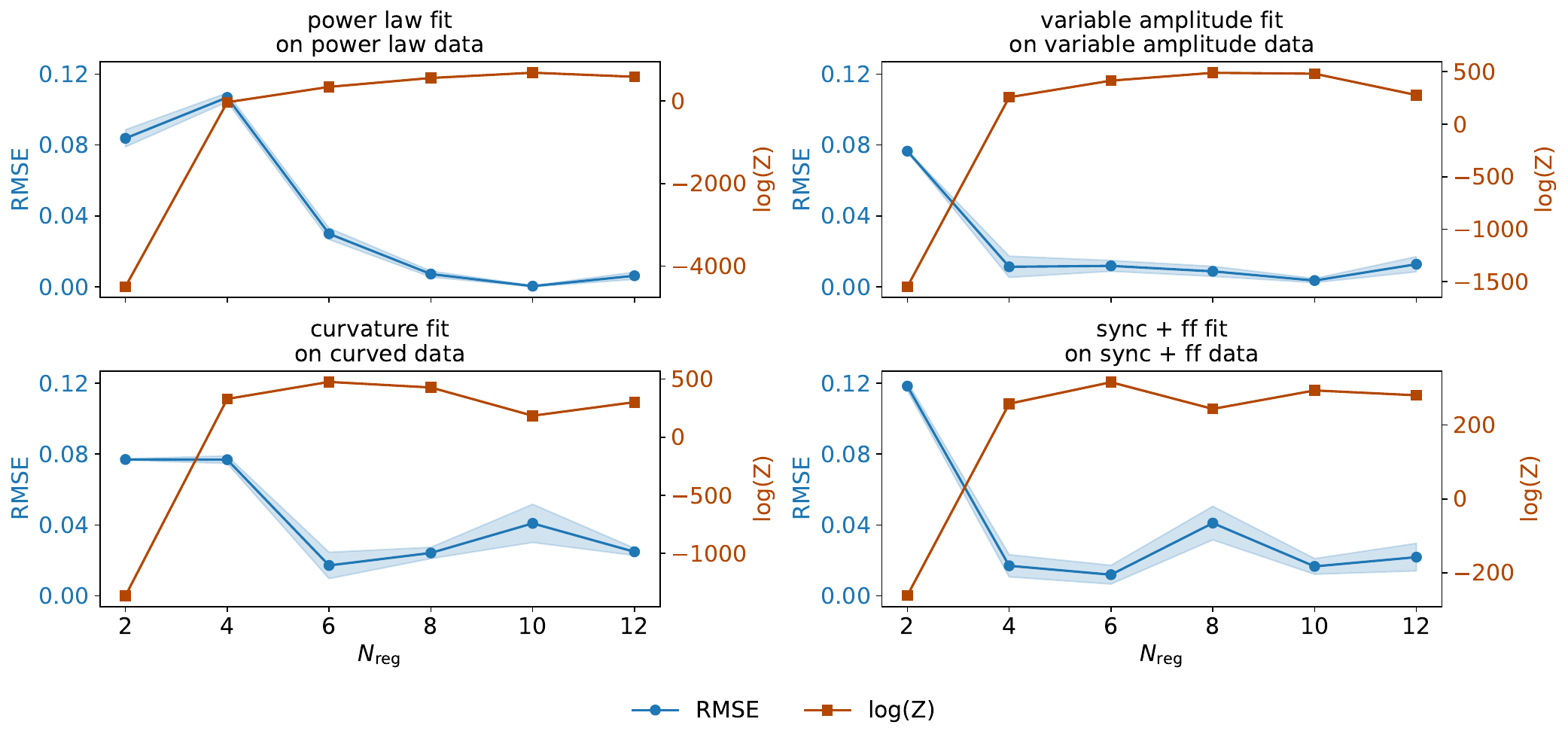}
\caption{Impact of increasing the number of regions on model performance.
Bayesian evidence and RMSE are shown as functions of $N_{\rm reg}$ for all four models fitted to their corresponding region-wise parameter data, using traditional (synchrotron-informed) region-splitting. }
\label{fig:nreg_comparison}
\end{figure*}

In order that an appropriate number of regions $N_{\rm reg}$ is used in these comparisons, we investigated how the value of $N_{\rm reg}$ affects the performance of each foreground model.
Figure~\ref{fig:nreg_comparison} shows Bayesian evidence and RMSE as functions of $N_{\rm reg}$ for all four models on their corresponding region-wise parameter data.
For the sake of computational feasibility, the number of live points was reduced (from $35 \times$ the number of dimensions used in all other runs to $25 \times$ the number of dimensions) in these tests.

Log-evidence rises, then approximately stabilises with $N_{\rm reg}$ for all models.
For the power law model, RMSE decreases as $N_{\rm reg}$ increases, reaching a minimum at $N_{\rm reg}=10$.
The variable amplitude and sync + ff models show similar RMSE saturation behaviour, with RMSE decreasing more rapidly than for the power law model and stabilising around $N_{\rm reg}\geq4$.
The curvature model, however, remains numerically unstable at the reduced live-point count used in these tests, and its RMSE does not reliably converge across the $N_{\rm reg}$ range explored.

Run time scales approximately as $N_{\rm dims}^3$ multiplied by the time complexity of the likelihood, defined in equation~(\ref{eq:ndims}) \citep{bevins2022marginal,lovick2026automatic}, so increasing $N_{\rm reg}$ has a significant computational cost.
The diminishing improvements in RMSE by increasing $N_{\rm reg}$ beyond 10 suggests that 10 regions capture the key spatial structure required for 21cm signal recovery and foreground mapping, so the choice of $N_{\rm reg}=10$ for our fiducial runs is justified as a good balance between model flexibility and computational efficiency.

\section{Results}
\label{sec:results}

We present results on three objectives: 21cm signal recovery (Section~\ref{sec:signal_extraction});  foreground parameter recovery (Section~\ref{sec:param_recovery}); and component-separated mapping with the sync + ff model (Section~\ref{sec:component_maps}). 

\subsection{21cm Signal Recovery}
\label{sec:signal_extraction}

\begin{figure*}
\includegraphics[width=0.989\textwidth]{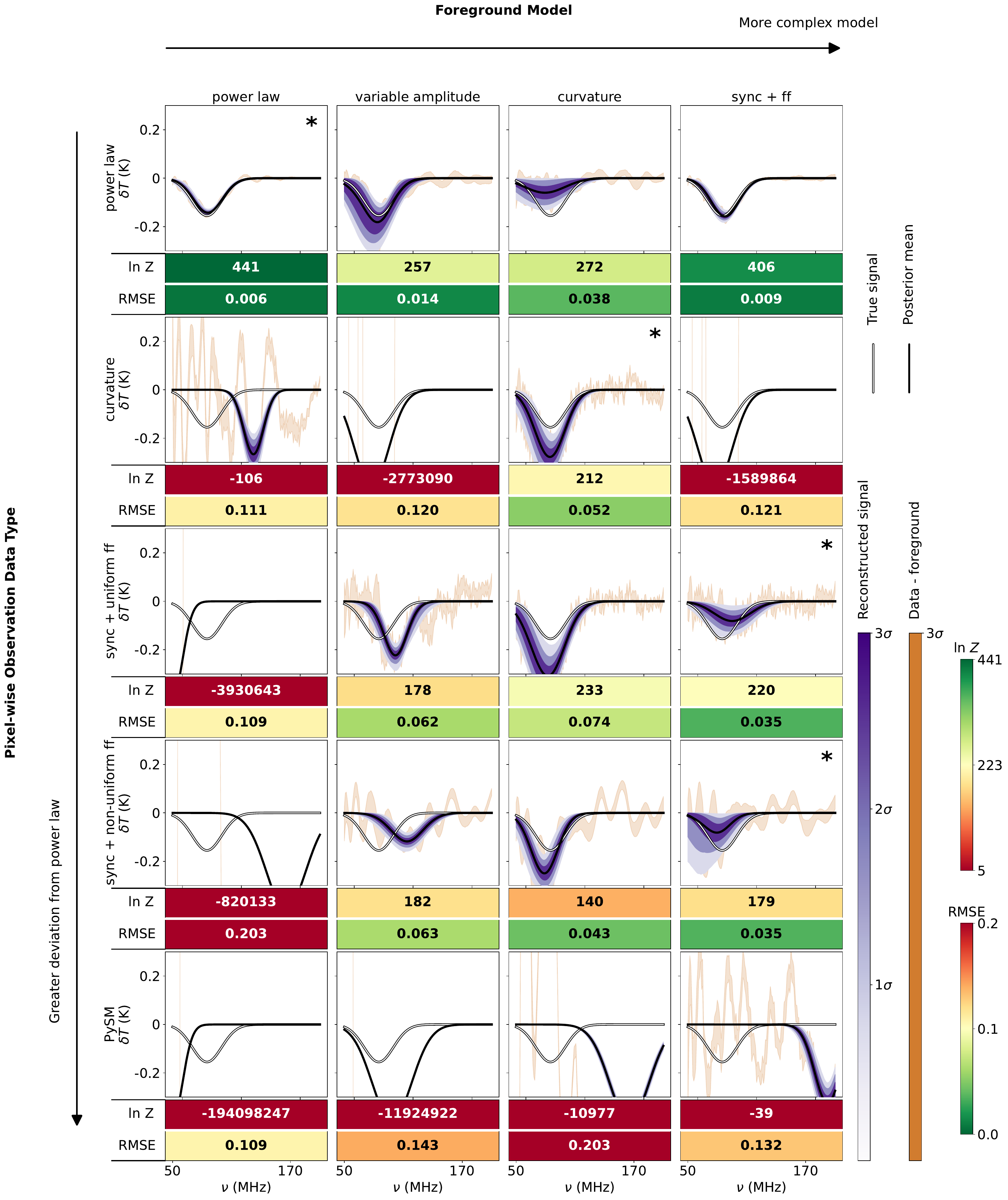}
\caption{A comparison of how well different foreground models recover the 21cm signal from different types of pixel-wise observation data.
Columns correspond to each of the four foreground models described in Section \ref{sec:models}, and rows correspond to each of the pixel-wise observation data types outlined in Section \ref{sec:simulation}.
In each main panel, the true injected 21cm signal is shown as a white line with a black outline.
The data with the fitted foreground removed ($\mathrm{data}-\mathrm{foreground}$) is shown by its 3$\sigma$ credible interval (orange contour). 
The reconstructed signal is shown by its posterior mean (solid black line) and its credible intervals (purple contours).
The inset statistics give the model evidence ($\ln \mathcal{Z}$) and signal-recovery error (RMSE).
The panels containing an asterisk (*) in the top-right corner mark the combinations where the foreground model matches the observation data type, and this is where we would expect a good signal recovery.
}
\label{fig:total_comparison_pixel-wise}
\end{figure*}

First, we assess 21cm signal recovery on pixel-wise data, summarised in Figure~\ref{fig:total_comparison_pixel-wise}.

\begin{figure*}
\includegraphics[width=\textwidth]{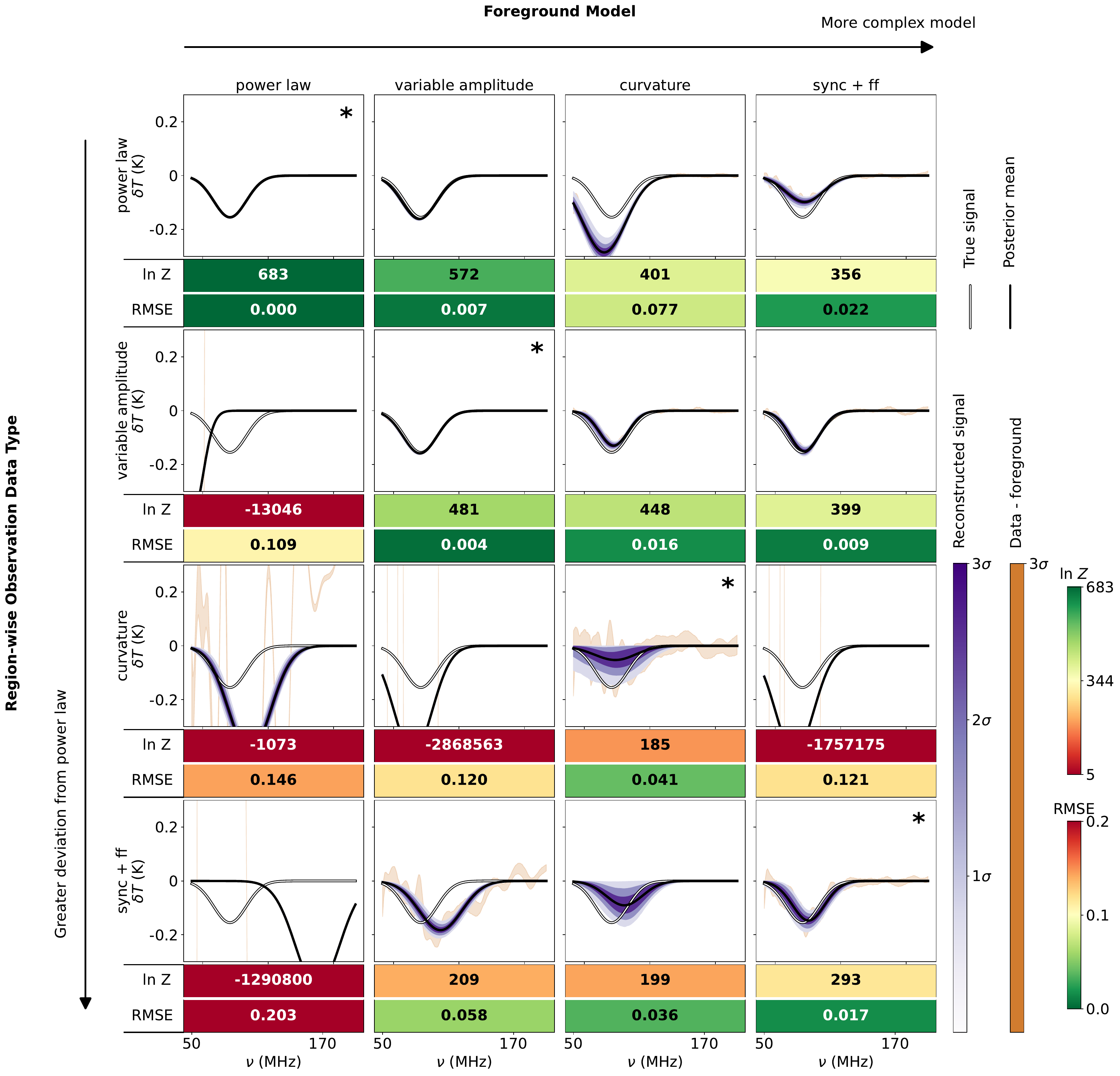}
\caption{A comparison of how well different foreground models recover the 21cm signal from different types of region-wise observation data.
Columns correspond to each of the four foreground models described in Section \ref{sec:models}, and rows correspond to each of the region-wise observation data types outlined in Section \ref{sec:simulation}.
In each main panel, the true injected 21cm signal is shown as a white line with a black outline.
The data with the fitted foreground removed ($\mathrm{data}-\mathrm{foreground}$) is shown by its 3$\sigma$ credible interval (orange contour). 
The reconstructed signal is shown by its posterior mean (solid black line) and its credible intervals (purple contours).
The inset statistics give the model evidence ($\ln \mathcal{Z}$) and signal-recovery error (RMSE).
The diagonal panels are marked by an asterisk (*); these are the combinations where the foreground model matches the observation data type, and this is where we would expect a good signal recovery.}
\label{fig:total_comparison_regionwise}
\end{figure*}

We highlight several features of Figure~\ref{fig:total_comparison_pixel-wise}.
First consider the left-hand column.
The simple power law model fits the 21cm signal very well when the data is a pure power law, but fails entirely as soon as the observation data deviates from this simple spectral structure.
The other three models are capable of fitting pure power law data, albeit with lower evidence, higher RMSE, and a wider credible interval.

In general, observation data types with richer spectral structure (in the lower rows) require more complex models (the right-hand columns) to successfully fit the 21cm signal.

However, the \texttt{PySM3} row shows poor signal recovery across all four models.
This is likely due to two factors.
Firstly, the \texttt{s1} synchrotron template is derived from a different observational basis than the GSM map used as the fitting anchor, meaning the spectral structure of \texttt{PySM3} is not well matched by our fitting models (this is also true of the sync + non-uniform ff data).
Secondly, small-scale spatial structure has been added directly to the \texttt{s1} template \citep{pan2025full}, so the \texttt{PySM3} sky contains finer spatial structure than our other data types, which are constructed at $1^\circ$ resolution; with only $N_{\rm reg} = 10$ regions, the model cannot resolve this structure.
Together, these lead to residual foreground structure after fitting, which corrupts the recovered cosmological signal.
The difficulty of separating synchrotron and free-free emission is well-known, and is discussed further in Section~\ref{sec:discussion}.

If the real sky at REACH frequencies exhibits spectral structure comparable in richness to \texttt{PySM3}, none of the models tested here will be sufficient to recover the 21cm signal without external constraints or a larger value of $N_{\rm reg}$.
It is worth noting that the \texttt{PySM3} sky is not necessarily more accurate than the other simulated observations at REACH frequencies, because it involved extrapolating well beyond the GHz frequencies from the surveys on which \texttt{PySM3} is calibrated.

Figure~\ref{fig:total_comparison_regionwise} shows the equivalent signal recovery grid for region-wise data.
As expected, signal recovery is generally better on these data compared with pixel-wise data.
The absence of pixel-level variations and beam effects in region-wise data leads to systematically lower RMSE values and tighter credible intervals compared to pixel-wise data.

Figure~\ref{fig:total_comparison_regionwise} has several notable features.
Firstly consider the diagonal, where the model matches the data type (e.g. a curvature model fitting curved data).
Each model performs best when applied to its matching data type, with typically $\Delta \log \mathcal{Z} > 10$ compared to mismatched models.

In both Figure \ref{fig:total_comparison_pixel-wise} and \ref{fig:total_comparison_regionwise}, the sync + ff model performs badly on curved power law data; this is because this foreground model does not include a curvature term.

\subsection{Foreground Parameter Recovery}
\label{sec:param_recovery}

\begin{figure*}
\includegraphics[width=\textwidth]{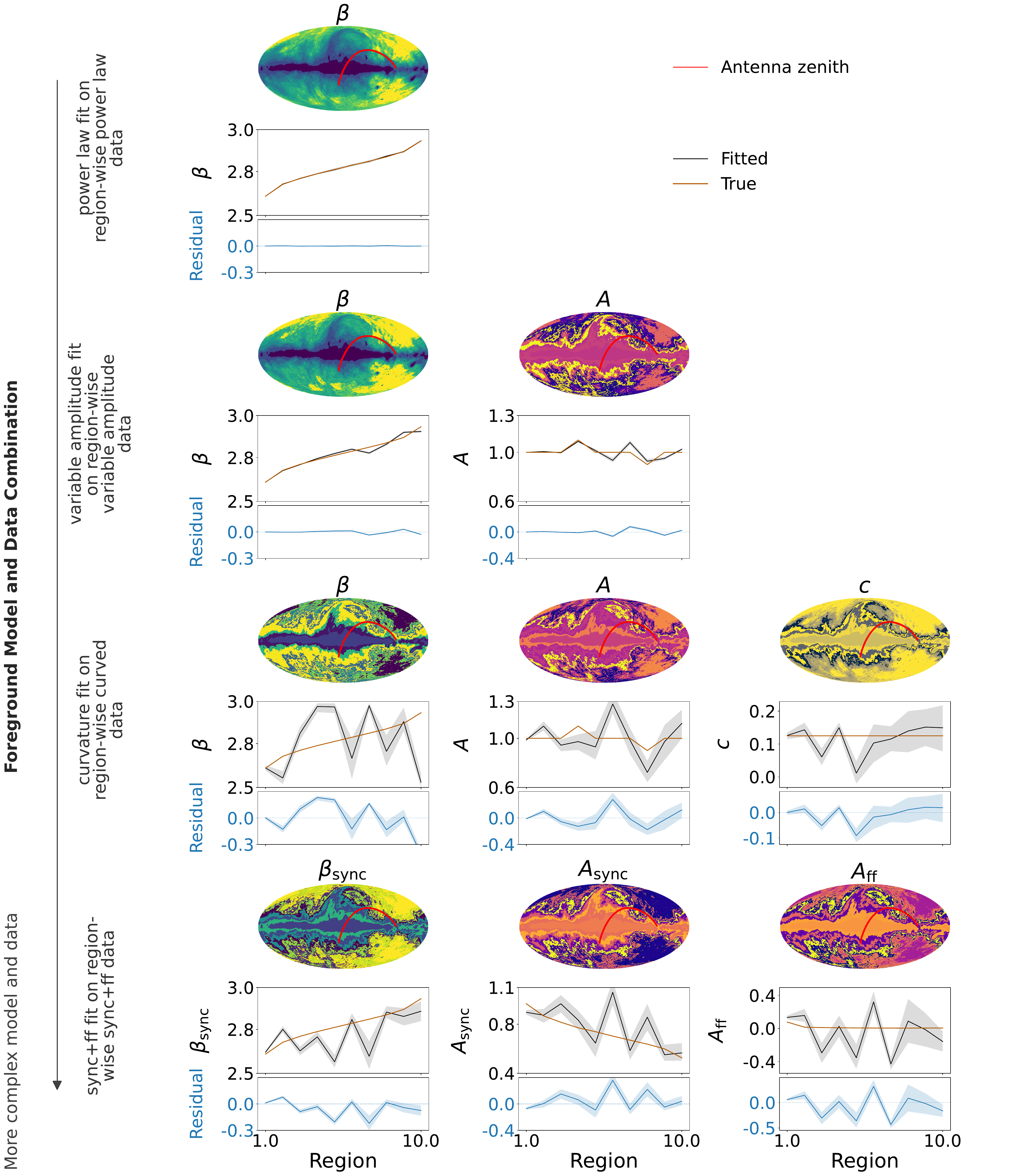}
\caption{Recovered-parameter summary for the four matching model / region-wise data combinations: power law fit on power law data (row 1), variable-amplitude power law fit on variable-amplitude data (row 2), curvature fit on curved data (row 3), and sync + ff fit on sync + ff data (row 4). In each row, the top panel shows the fitted parameter sky map, the middle panel shows fitted (black) vs true (dark orange) region-wise parameter values, and the bottom panel shows residuals (fitted minus true, blue) with 95\% credible interval shading propagated from the nested-sampling posterior samples. The red line on each sky map indicates the 12-hour LST track.}
\label{fig:diagonal_recovery}
\end{figure*}

To assess foreground parameter recovery, we compare the fitted vs true foreground parameters for the cases where the foreground model matches the region-wise observation data type (i.e. for the combinations along the diagonal of Figure \ref{fig:total_comparison_regionwise}).
This is shown in Figure~\ref{fig:diagonal_recovery}.
Each row corresponds to one model-data pair where the functional form matches the type of observation data (e.g. in the top row, the power law model on power law data).
Within each row, the top panel shows the fitted parameter sky map, the middle panel compares fitted vs true parameter values, and the bottom panel shows the residual (fitted minus true).
(Note that these parameter maps are the same at every frequency; it's only $T_{\rm sky}$ that is frequency-dependent.)

The top row of Figure \ref{fig:diagonal_recovery} represents the simplest model on the simplest data, fitting just $N_{\rm reg}$ parameters.
Correspondingly, this achieves excellent recovery with all fitted values consistent with the true values within the credible intervals. 

The second row, with the more complex variable amplitude model, which fits $2N_{\rm reg}$ parameters, shows generally good parameter recovery for both $A$ and $\beta$, but not as good as in the pure power law case.

Moving to the bottom two rows, where the models fit $3N_{\rm reg}$ parameters, we see poor recovery of the true parameters.
This is attributable to the strong degeneracy between the parameters, a fundamental issue in this field (see \citealp{hibbard2023fitting}; \citealp{irfan2026measuring}; and Section~\ref{sec:discussion}).
Additional parameters increase the effect of this degeneracy, so foreground modelling for 21cm cosmology should optimise model complexity such that it is sufficient to fit the foreground spectral structure without unnecessarily amplifying degeneracy effects.

The physical origin of these degeneracies is worth understanding: in the curvature model, a steeper spectral index $\beta$ can mimic positive curvature $c$ over a limited frequency baseline, because both have the effect of reducing emission at higher frequencies relative to lower ones.
In the sync + ff model, a larger synchrotron amplitude $A_{\rm sync}$ can partially absorb what would otherwise be attributed to free-free emission, particularly in sky regions where the two components have similar spectral shapes over the REACH band.
So, even with good signal-to-noise, individual foreground parameters remain degenerate unless informative priors or additional external constraints (e.g.\ from new surveys) are used to break the degeneracy.

It is worth noting, in Figure \ref{fig:diagonal_recovery}, that all models fit more successfully at the extreme regions: very near to the Galactic plane (where the sky temperature is highest, and therefore the signal-to-noise is best) and very far from the Galactic plane (where few individual radio sources affect the region boundaries), compared with the intermediate regions (where many individual sources affect the region boundaries - see Figure \ref{fig:region_latitude_explainer}).
However, the outer regions contribute less to the total antenna temperature, so their poorer parameter recovery has a smaller impact on both the recovered foreground and signal.
The degeneracy structure is quantified further in Sections~\ref{sec:degeneracies} and~\ref{sec:corr_coefficients}.

\subsubsection{Parameter degeneracies}
\label{sec:degeneracies}

Understanding correlations between fitted parameters is crucial for assessing model reliability and identifying how degeneracies affect physical interpretation.
Degeneracies exist between individual foreground components, and also between the foreground and 21cm signal \citep{hibbard2023fitting}.

We examine parameter degeneracies for the three multi-parameter models.
For a model with $p_{\rm model}$ parameters, there are $\binom{p_{\rm model}}{2} = p_{\rm model}(p_{\rm model}-1)/2$ possible parameter pairs, so the variable amplitude model has one pair, and the curvature and sync + ff models have three pairs each.
We consider only pairs of parameters within the same region; cross-region correlations are also possible, since all regions contribute to a single observed spectrum, but these correlations are not considered here.

\begin{figure*}
\includegraphics[width=\textwidth]{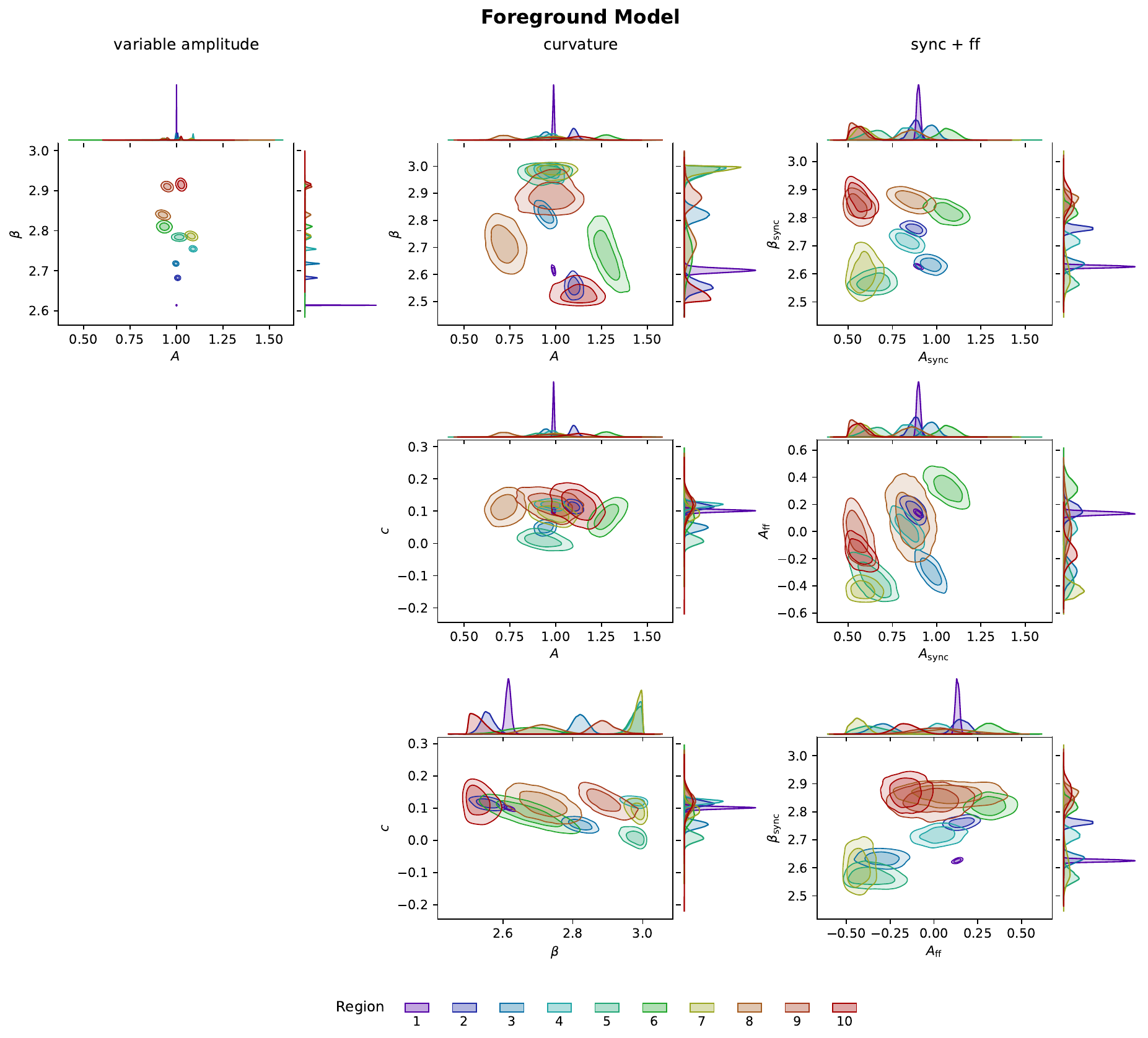}
\caption{Parameter degeneracies for the three multi-parameter models, with all 10 regions plotted simultaneously (different colours/contours per region). Columns correspond to the variable amplitude power law (left), curved power law (centre), and sync + ff (right) models. 
The variable amplitude power law column shows one pair plot ($A$ vs $\beta$); the curved power law column shows three pair plots ($A$ vs $\beta$, $A$ vs $c$, and $\beta$ vs $c$); and the sync + ff column shows three pair plots ($A_{\rm sync}$ vs $\beta_{\rm sync}$, $A_{\rm sync}$ vs $A_{\rm ff}$, and $A_{\rm ff}$ vs $\beta_{\rm sync}$).
Running along the top and right edges of each panel are the posterior distributions of the fitted parameters in each region.
}
\label{fig:model_degeneracies}
\end{figure*}

Figure~\ref{fig:model_degeneracies} shows the joint posterior distributions for all parameter pairs across all regions, for these three models on their corresponding observation data type. The three columns correspond to the variable amplitude power law, curved power law, and sync + ff models, again on their corresponding region-wise observation data.

Degeneracies appear as elongated diagonal contours.
In the variable amplitude column, the degeneracies are weak and $A$ and $\beta$ are well-constrained.
$A$ shows minimal correlation with $\beta$ across all regions, reflecting the fact that $A$ acts as a spatial multiplicative scaling that does not affect spectral shape. This ensures that brightness normalisation uncertainties do not significantly impact spectral characterisation.

However, the more complex models show stronger degeneracies and poorer constraints on the parameters.
In the curved power law model, $A$ shows minimal correlation with $c$ across all regions, but there is a negative correlation between $\beta$ and $c$ in most regions. This is physically expected: steeper spectral indices (more negative $\beta$) can partially compensate for positive curvature over limited frequency ranges, and vice versa. 

In the sync + ff model, there is also some trade-off between $A_{\rm sync}$ and $A_{\rm ff}$, which limits the viability of mapping separated synchrotron and free-free components.
$A_{\rm sync}$ and $\beta_{\rm sync}$ show modest mutual correlations similar to those seen in the curved power law. 
$A_{\rm ff}$ shows broader posteriors and larger uncertainties than the synchrotron parameters, reflecting the subdominant free-free contribution. 

Figure \ref{fig:model_degeneracies} also shows some region-dependent structure for all models: different regions show varying degrees of parameter correlation; regions with stronger foreground emission (especially Region 1, covering the Galactic plane) generally show tighter posteriors and weaker correlations.
This reflects the fact that outer regions contribute less to the total sky power, so the associated foreground parameters are less constrained.

The correlations observed are significant but not absolute: the posteriors are elongated but not perfectly degenerate, so additional parameters still provide independent information.
If parameters were fully degenerate, we would observe narrow diagonal ridges in the joint posteriors rather than the relatively broad distributions seen here.
This validates the use of multi-parameter models for extracting physical information from foreground observations.

\subsubsection{Posterior linear correlation coefficients}
\label{sec:corr_coefficients}

To quantify the degeneracy structure seen in Figure~\ref{fig:model_degeneracies} more precisely, we compute the posterior linear correlation coefficient $\rho_{xy}$ between key parameter pairs as a function of sky region $i$ for each model.
For weighted posterior samples $(x_i, y_i, w_i)$ with $\sum_i w_i = 1$, this is defined as
\begin{equation}
\rho_{xy} = \frac{\sum_i w_i (x_i - \mu_x)(y_i - \mu_y)}{\sigma_x \sigma_y},
\end{equation}
where $\mu_x = \sum_i w_i x_i$ and $\sigma_x^2 = \sum_i w_i (x_i - \mu_x)^2$ (and analogously for $y$).

\begin{figure*}
\includegraphics[width=\textwidth]{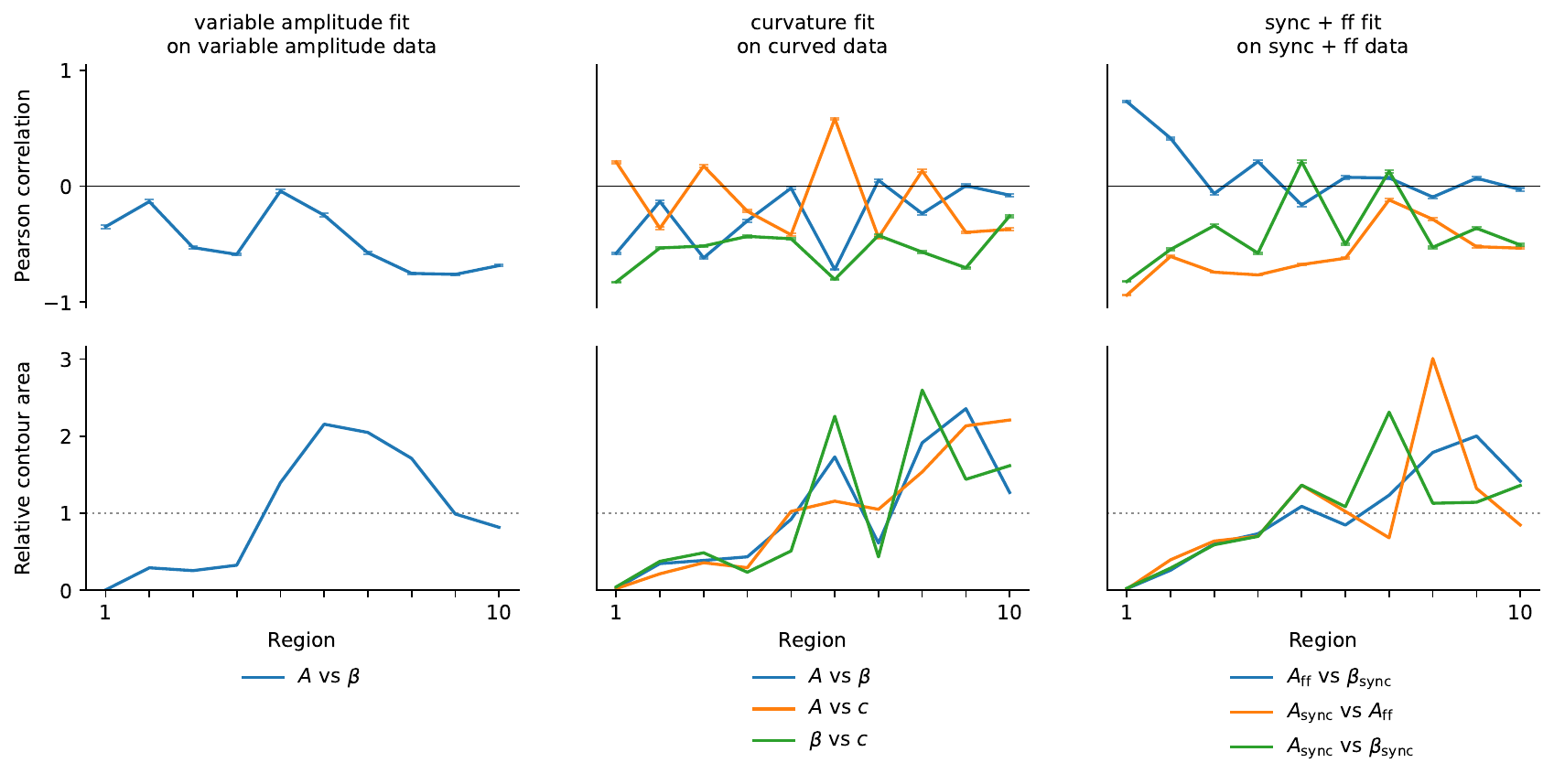}
\caption{Region-by-region posterior degeneracy structure for key parameter pairs in the three multi-parameter models (variable-amplitude power law (left), curved power law (middle), and sync + ff (right), each fitted to their corresponding observation data type.
\textit{Top row:} Weighted Pearson correlation coefficient $\rho_{xy}$ as a function of sky region. 
The error bars show the 1$\sigma$ uncertainty.
\textit{Bottom row:} Relative posterior area $A_i^{\rm rel}$, normalised to the pair-specific mean across regions. Values below unity indicate tighter-than-typical constraints; values above unity indicate broader-than-typical constraints.
Reading both rows together separates degeneracy shape (tilt, captured by $\rho_{xy}$) from constraint strength (contour size, captured by $A_i^{\rm rel}$).
The grey dotted line indicates relative contour area = 1, above which the constraint on a given parameter pair is broader (i.e. poorer) than average.}
\label{fig:corr_coefficients}
\end{figure*}

The upper row of Figure~\ref{fig:corr_coefficients} shows Pearson correlation $\rho_{xy}$ as a function of region for all parameter pairs within each of the three multi-parameter models.
The strongest anti-correlations are $A_{\rm sync}$-$A_{\rm ff}$ in the sync + ff model (mean $\rho \approx -0.76$) and $\beta$-$c$ in the curvature model (mean $\rho \approx -0.70$), reflecting fundamental parameter trading: a larger synchrotron amplitude is partially offset by a smaller free-free amplitude, and a steeper spectral index partially compensates for positive curvature.
The $A$-$\beta$ pair is moderately anti-correlated in both the variable amplitude and curvature models (means $\approx -0.49$ and $-0.46$ respectively).
By contrast, $A$-$c$ and $A_{\rm ff}$-$\beta_{\rm sync}$ are weaker and change sign across regions, indicating region-dependent posterior geometry rather than a universal degeneracy.

Because correlation is scale-free, a large $|\rho_{xy}|$ can occur even when both parameters are individually well-constrained.
To separate degeneracy \emph{shape} from constraint \emph{strength}, the bottom row of Figure~\ref{fig:corr_coefficients} shows the relative posterior area $A_i^{\rm rel}$ for the same parameter pairs and regions.
For a given parameter pair $(x,y)$ in region $i$, we construct the weighted covariance matrix
\begin{equation}
\Sigma_{xy\mathit{i}} =
\begin{pmatrix}
\sigma_{x\mathit{i}}^2 & \mathrm{Cov}_{xy\mathit{i}}\\
\mathrm{Cov}_{xy\mathit{i}} & \sigma_{y\mathit{i}}^2
\end{pmatrix},
\end{equation}
with weighted moments evaluated from posterior samples (weights normalised to $\sum_j w_j = 1$). An ellipse-area proxy is then defined as
\begin{equation}
A_i \propto \sqrt{\det(\Sigma_{xy\mathit{i}})} = \sqrt{\sigma_{x\mathit{i}}^2\,\sigma_{y\mathit{i}}^2 - \mathrm{Cov}_{xy\mathit{i}}^2},
\end{equation}
and normalised within each parameter pair by the median area across regions:
\begin{equation}
A_i^{\rm rel} = \frac{A_i}{\mathrm{median}_{i}(A_{i})}.
\end{equation}
Values $A_i^{\rm rel} < 1$ indicate tighter-than-typical constraints for that pair in that region; values $A_i^{\rm rel} > 1$ indicate broader-than-typical constraints.
Inspecting both rows of Figure~\ref{fig:corr_coefficients} together therefore separates regions where a high correlation reflects a genuine physical degeneracy from those where the posterior contour is simply small, reducing the risk of over-interpreting large $|\rho_{xy}|$ values in well-constrained regions.

The bottom row shows that $A_i^{\rm rel}$ generally increases from low- to high-numbered regions, which correspond to progressively lower sky temperatures away from the Galactic plane (see Figure~\ref{fig:region_latitude_explainer}).
This is physically expected: fainter regions have a lower signal-to-noise ratio and contribute a smaller fraction of the total sky temperature, yielding looser constraints on all parameters.
The curvature model (middle column) shows the most variation in $A_i^{\rm rel}$ across regions, with the $\beta$-$c$ pair displaying especially large areas at high region numbers.
This is consistent with the recurring difficulty of constraining $c$ in low-brightness regions.

\subsubsection{Sky map reconstruction validation}

\begin{figure*}
\includegraphics[width=\textwidth]{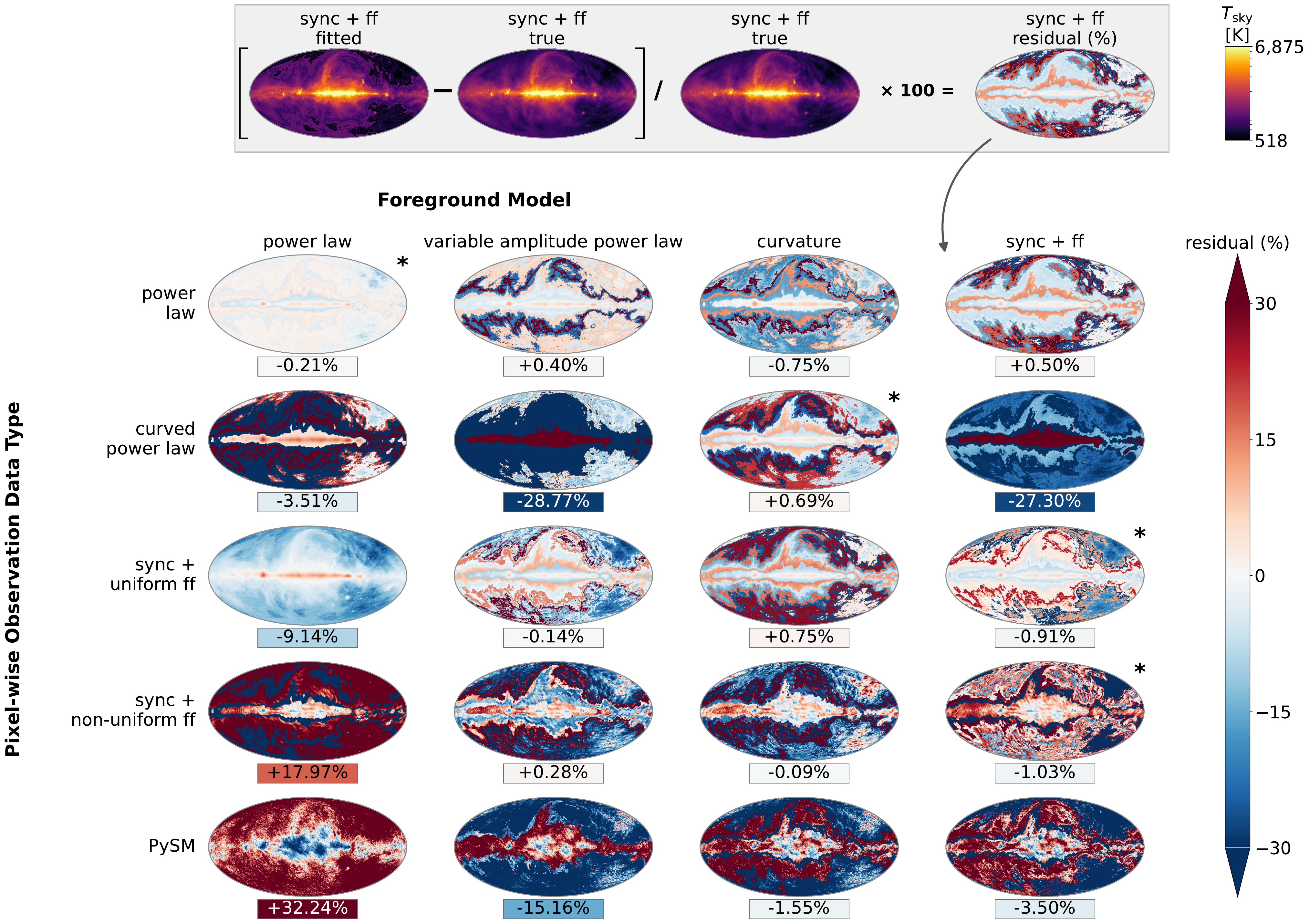}
\caption{Sky map reconstruction at 100 MHz for all four foreground models applied to pixel-wise observation data.
Columns correspond to the four foreground models, and rows correspond to the five pixel-wise observation data types.
For each model/data combination, the percentage residual, defined as $(T_{\rm sky,fitted} - T_{\rm sky,true})/T_{\rm sky,true} \times 100$\% is shown.
Under each map, the mean error across the whole sky is shown, using the same colour scale as the error for each region.
The maps marked with an asterisk correspond to the `matched' data/model combinations.
The shaded inset at the top illustrates this residual definition using the sync + ff fit as an example, with the corresponding true, fitted, and residual maps shown.}
\label{fig:mastermap_continuous}
\end{figure*}

\begin{figure*}
\includegraphics[width=\textwidth]{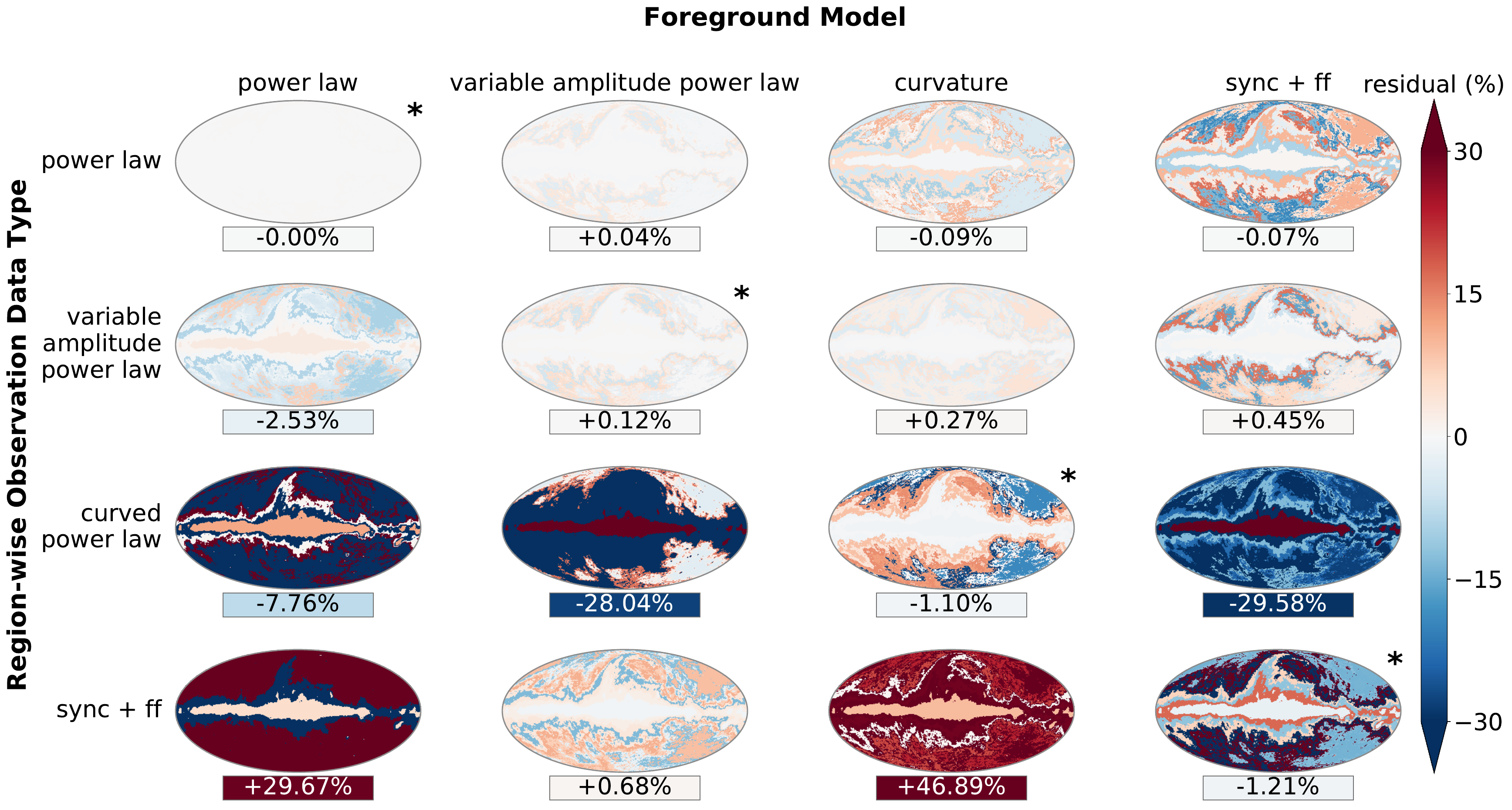}
\caption{Sky map reconstruction at 100 MHz for all four foreground models applied to region-wise observation data.
Columns correspond to the four foreground models, and rows correspond to the five region-wise observation data types.
For each model/data combination, the percentage residual, defined as $(T_{\rm sky,fitted} - T_{\rm sky,true})/T_{\rm sky,true} \times 100$\% is shown.
Under each map, the mean error across the whole sky is shown, using the same colour scale as the error for each region.
The maps marked with an asterisk correspond to the `matched' data/model combinations.
}
\label{fig:mastermap_discrete}
\end{figure*}

Beyond parameter recovery, we assess how well each model reconstructs the full sky temperature distribution.
Even when parameters are degenerate, they should still combine to give the true sky temperature.
Each combination of foreground model and observation data type gives fitted sky temperature maps at continuous frequencies between 50-170\,MHz; here, we show only the percentage residual between the fitted maps and the true map (from Section \ref{sec:simulation}) at 100\,MHz.
We show this for pixel-wise observation data in Figure~\ref{fig:mastermap_continuous} and for region-wise data in Figure~\ref{fig:mastermap_discrete}.

Again, the combinations where the foreground model matches the observation data type show the tightest residuals.
Region-wise data yield cleaner reconstructions than pixel-wise data, reflecting the absence of pixel-level variations and beam convolution effects.

The radiometric noise injected into the data simulations also accounts for some of the discrepancy between true and fitted parameter values.

\subsection{Component-Separated Foreground Maps with the Synchrotron + free-free Model}
\label{sec:component_maps}

Component-separated mapping is a capability unique to the sync + ff foreground model, provided that foreground parameters are recovered accurately.
The sync + ff model enables construction of separate synchrotron and free-free emission maps using the fitted amplitudes and spectral indices:
\begin{equation}
T_{\rm sync,\mathit{i}}(\Omega,\nu) = A_{\rm sync,\mathit{i}}[T_{230}(\Omega) - T_{\rm CMB}]\left(\frac{\nu}{230~{\rm MHz}}\right)^{-\beta_{\rm sync,\mathit{i}}} + T_{\rm CMB},
\end{equation}
\begin{equation}
T_{\rm ff,\mathit{i}}(\Omega,\nu) = A_{\rm ff,\mathit{i}}[T_{230}(\Omega) - T_{\rm CMB}]\left(\frac{\nu}{230~{\rm MHz}}\right)^{-2.1} + T_{\rm CMB}.
\end{equation}
To suppress region-boundary artefacts arising from any single $N_{\rm reg}$ choice, we combine maps from multiple runs using evidence-weighted averaging, described in Section~\ref{sec:map_averaging} below.

\subsubsection{Evidence-weighted map averaging}
\label{sec:map_averaging}

A single $N_{\rm reg}$ run produces region-wise constant-parameter fitted maps with discontinuities at region boundaries.
To suppress these artefacts, we combine fits from multiple runs spanning $N_{\rm reg} \in \{2, 4, 6, 8, 10, 12\}$ using an evidence-weighted average.
Each fit $i$ contributes according to its Bayesian evidence $\mathcal{Z}_i$, with normalized weights
\begin{equation}
w_i = \frac{\exp\!\left(\log \mathcal{Z}_i - \max_k \log \mathcal{Z}_k\right)}
           {\sum_{k} \exp\!\left(\log \mathcal{Z}_k - \max_k \log \mathcal{Z}_k\right)},
\label{eq:evidence_weights}
\end{equation}
which is equivalent to $w_i \propto \mathcal{Z}_i$ with $\sum_i w_i = 1$.
The numerically stable form in equation~(\ref{eq:evidence_weights}) subtracts the maximum log-evidence before exponentiating to avoid floating-point underflow when log-evidences are large and negative.
Note that if one run has substantially higher evidence than the others, its weight will dominate, and the averaged map approaches the single best-fit map.

For each pixel $p$, the evidence-weighted component maps are
\begin{equation}
\bar{T}_{\rm ff}(p) = \sum_i w_i\,T_{{\rm ff}\mathit{i}}(p), \qquad
\bar{T}_{\rm sync}(p) = \sum_i w_i\,T_{{\rm sync}\mathit{i}}(p),
\label{eq:weighted_components}
\end{equation}
from which the overall map is obtained as
\begin{equation}
\bar{T}_{\rm total}(p) = \bar{T}_{\rm ff}(p) + \bar{T}_{\rm sync}(p).
\end{equation}
Because the averaging is performed in map space (pixel by pixel) rather than in parameter space, it naturally smooths region-boundary discontinuities in $\bar{T}_{\rm total}(p)$: pixels near a boundary receive contributions from multiple runs whose boundaries lie at different locations, producing a gradual transition rather than a sharp step.
The evidence weighting ensures that better-fitting $N_{\rm reg}$ values contribute more to the final map, which also reduces the residuals compared to a single $N_{\rm reg}$ run.

\begin{figure*}
\includegraphics[width=0.95\textwidth]{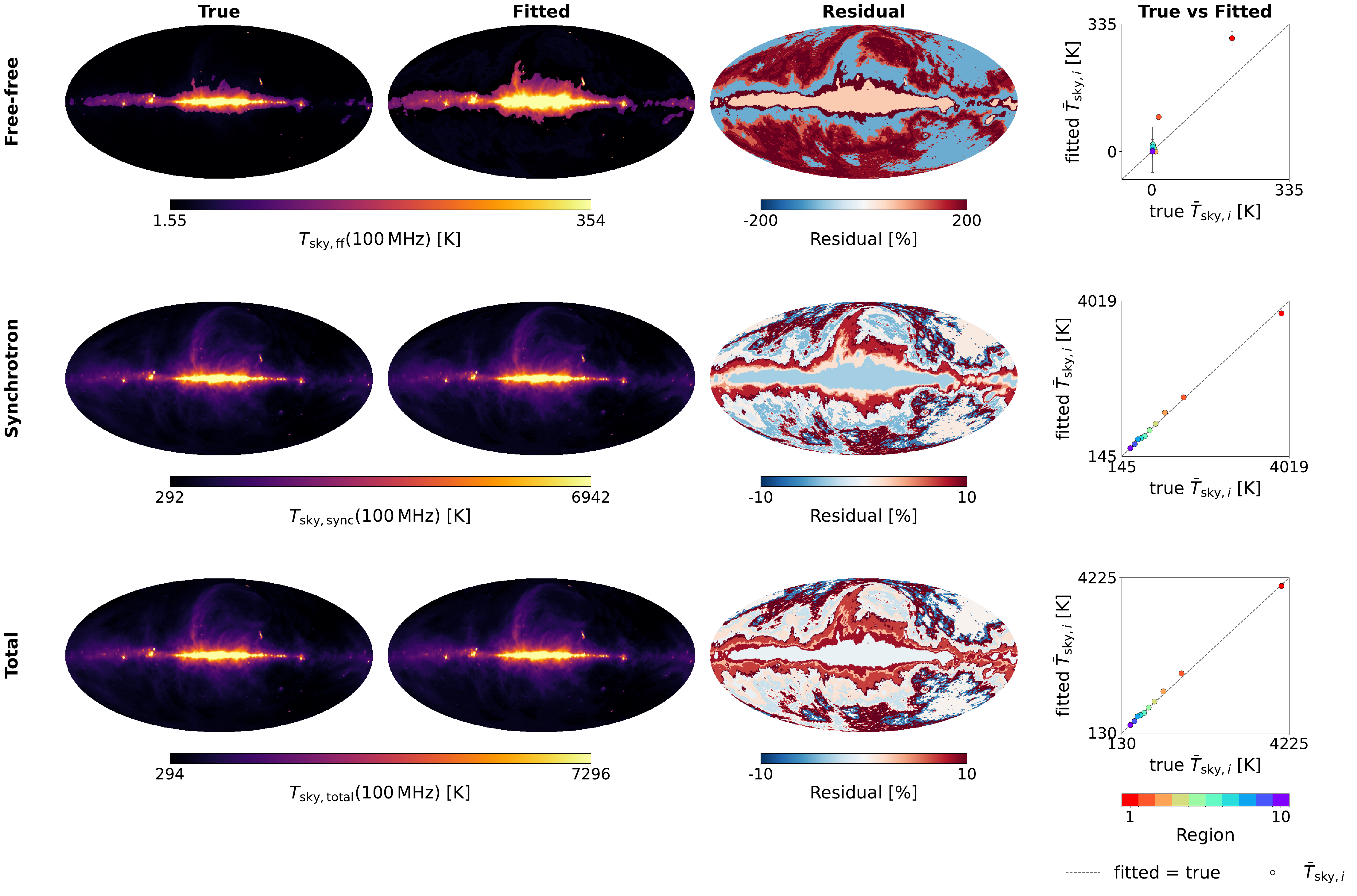}
\caption{True, evidence-weighted fitted, and percentage residual maps at 100 MHz for the sync + ff component separation, averaged across runs with $N_{\rm reg} \in \{2, 4, 6, 8, 10, 12\}$ using the Bayesian evidence weights defined by equation~(\ref{eq:evidence_weights}).
This is the sync + ff fit applied to region-wise sync + ff data.
Rows show free-free emission (top), synchrotron emission (middle), and total (synchrotron + free-free) emission (bottom).
Columns show the true map (left), evidence-weighted fitted map (centre), and percentage residual (right).
The free-free residual colour scale is capped at $\pm200 \%$: because free-free temperatures are very low, a small absolute error produces a large fractional residual.
Synchrotron and total residuals are capped at $\pm10$\,per\,cent.
}
\label{fig:syncff_component_maps}
\end{figure*}

Figure~\ref{fig:syncff_component_maps} shows the resulting evidence-weighted component maps at 100~MHz, for the region-wise sync + ff observation data.

Residuals for the synchrotron and total emission maps are generally within $\pm10\%$ across the sky, indicating that the dominant emission component is well recovered without systematic error.
However, the free-free component shows larger fractional residuals, due to the very low free-free brightness temperatures. 
Remaining systematic errors in the free-free component arise from the mismatch between fitted region geometry and the physical distribution of free-free emission: in reality, free-free emission is concentrated in a thin band along the Galactic plane, but the innermost fitted region spans a considerably thicker latitude range.
Bright compact radio sources may also bias the fitted free-free amplitude parameters in their host region: because all pixels in a region share the same fitted parameters, the source biases the fit and introduces systematic errors across the other pixels in that region (see Figure~\ref{fig:region_latitude_explainer}).

The fitted free-free map in Figure \ref{fig:syncff_component_maps} shows that synchrotron emission from the North Polar Spur is `leaking' into the free-free map, even though negligible free-free emission exists in this area.
This limitation persists even after evidence-weighted averaging, as it is geometric rather than statistical in origin, but it does diminish as $N_{\rm reg}$ increases.
Another method of splitting the sky is therefore required for the sync + ff model to accurately fit the free-free emission.
This splitting should be informed by known synchrotron and free-free emission maps, and is demonstrated in the following section.

\subsubsection{Free-free-informed and mixed region-splitting strategies}
\label{sec:splitting}

The large residuals in the free-free component identified above arise because the standard percentile-splitting method groups pixels near the Galactic plane together with higher-latitude pixels in the same innermost region.
Since free-free emission is concentrated in a thin band along the Galactic plane, a single region-averaged amplitude cannot simultaneously describe the bright pixels within the thin band and the more diffuse emission at higher latitudes within the same region.
We therefore introduce two additional region-partition strategies and assess their impact on foreground recovery.

For all three methods, the sky is partitioned into $N_{\rm reg} = 10$ regions with binary masks $M_i(p) \in \{0,1\}$ ($i = 1,\dots,10$) satisfying $\sum_i M_i(p) = 1$ for every HEALPix pixel $p$.

Traditional (synchrotron-informed) splitting is the percentile split of the spectral index map described in Section~\ref{sec:models}, which is dominated by synchrotron emission.

In FF-informed splitting, the partition is derived instead from the free-free brightness template $S_{\rm ff}(p)$ at 230\,MHz.
This map is computed from the emission measure map of \citet{Hutschenreuter2024} and the electron temperature map from the Planck Commander free-free product \citep{adam2016planck}, using the same Gaunt-factor optical-depth model described in Section~\ref{sec:simulation}; it therefore encodes the physical free-free brightness distribution rather than the synchrotron spectral structure used by the traditional split.
Pixels are ranked in descending order of $S_{\rm ff}$, and the sorted pixel list is split into $N_{\rm reg} = 10$ equal-count regions.
Region 1 therefore contains the $10\%$ of pixels with the highest free-free brightness, which are concentrated in the Galactic plane, while region 10 contains the faintest $10\%$ of pixels.

The mixed splitting strategy uses two independent region systems simultaneously: the traditional masks $M_i^{\rm sync}$ for the synchrotron parameters, and the FF-informed masks $M_i^{\rm ff}$ for the free-free amplitude.
The sky model is therefore
\begin{align}
A_{\rm sync}(p) &= \sum_i A_{{\rm sync}\mathit{i}}\,M_i^{\rm sync}(p), \\
\beta_{\rm sync}(p) &= \sum_i \beta_{{\rm sync}\mathit{i}}\,M_i^{\rm sync}(p), \\
A_{\rm ff}(p) &= \sum_i A_{{\rm ff}\mathit{i}}\,M_i^{\rm ff}(p).
\end{align}
This allows the synchrotron parameters to vary with the spectral-index structure of the sky, while ensuring that the free-free amplitude regions are aligned with the physical distribution of free-free emission.
The number of free parameters remains $3N_{\rm reg}$, unchanged from the standard sync + ff model.

\begin{figure}
\includegraphics[width=\columnwidth]{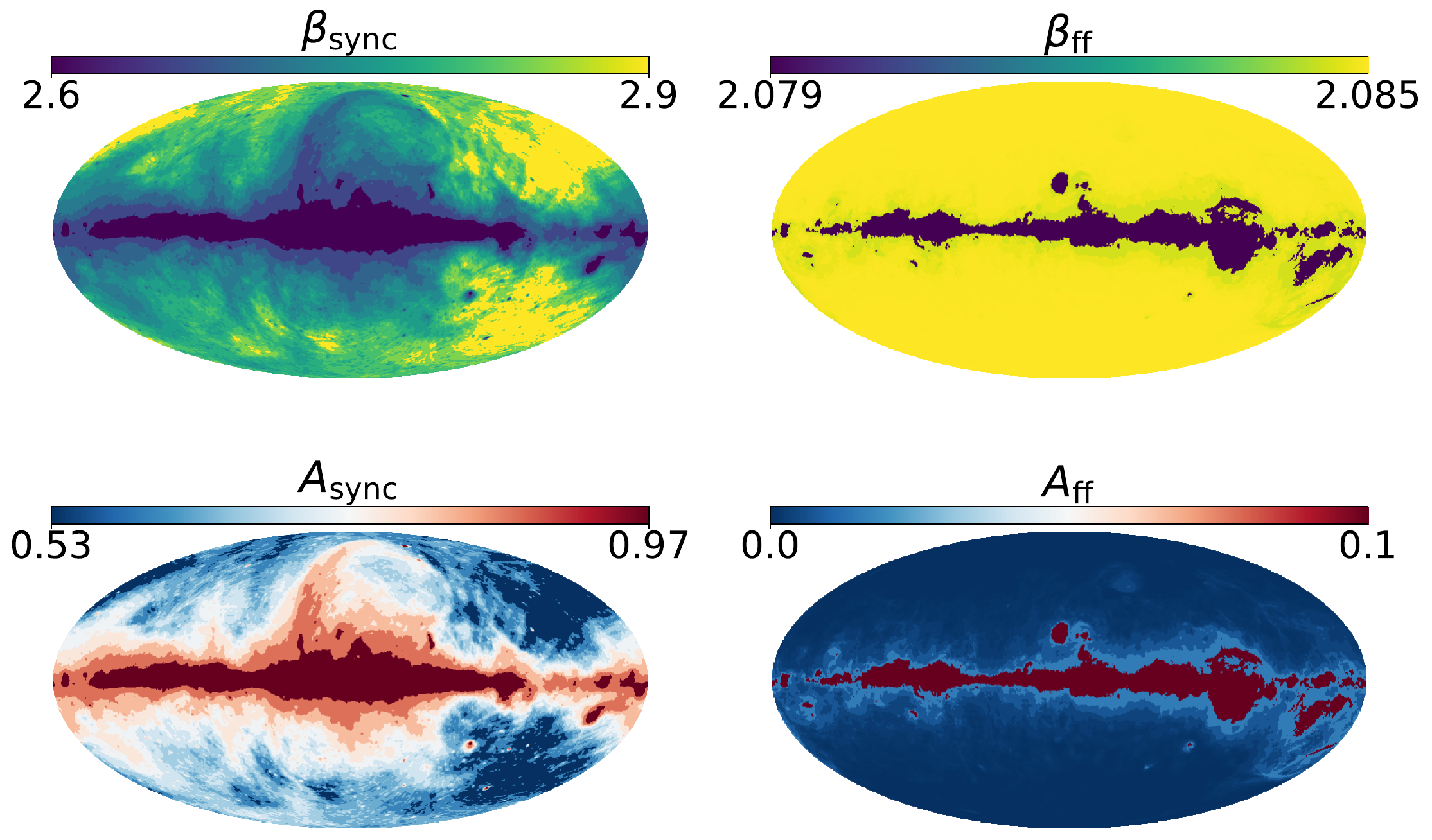}
\caption{Input parameter maps for the mixed-split dataset.
The synchrotron spectral index $\beta_{\rm sync}$ (top left) and amplitude $A_{\rm sync}$ (bottom left) are defined using the traditional synchrotron-informed partition, while the free-free spectral index (top right) and amplitude $A_{\rm ff}$ (bottom right) are defined on the FF-informed partition.
In both cases, $N_{\rm reg} = 10$.}
\label{fig:mixed_region_maps}
\end{figure}

To test these strategies under controlled conditions, we generate a new matched dataset using the mixed-split geometry: the true parameters $A_{{\rm sync}\mathit{i}}$, $\beta_{{\rm sync}\mathit{i}}$, and $A_{{\rm ff}\mathit{i}}$ are assigned on their respective synchrotron-informed and FF-informed region partitions, and observations are simulated from this truth using the standard pipeline described in Section~\ref{sec:simulation}.
The mixed splitting can be applied to any of the existing datasets, but this demonstration is designed to test the accuracy of parameter recovery, hence only `matched' data is shown in full.
Figure~\ref{fig:mixed_region_maps} shows the four input parameter maps ($A_{\rm sync}$, $\beta_{\rm sync}$, $A_{\rm ff}$, $\beta_{\rm ff}$) for this mixed-split dataset.

\begin{figure*}
\includegraphics[width=\textwidth]{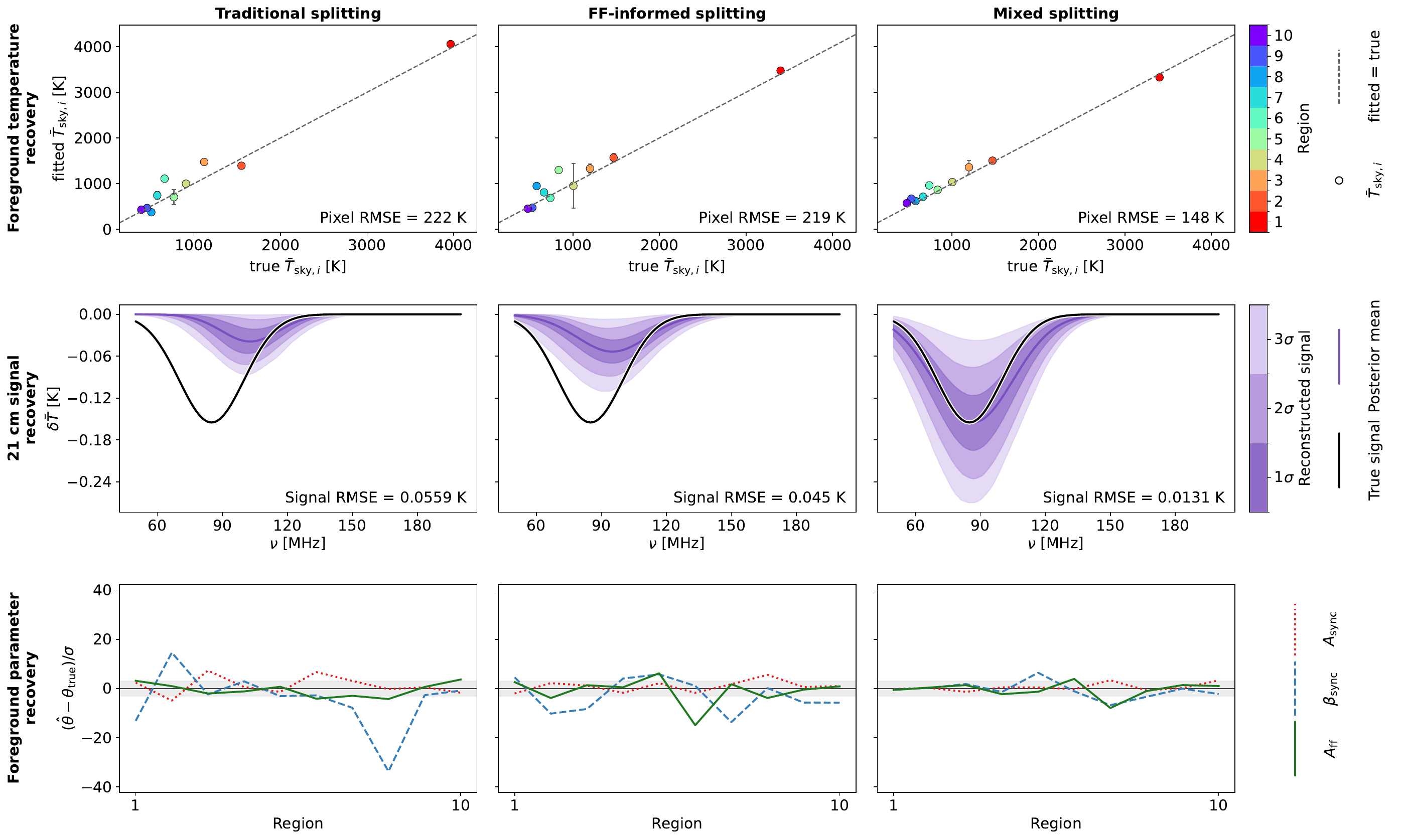}
\caption{Comparison of recovery performance for the $N_{\rm reg}=10$ 12-hour integrated mixed-split dataset, using the sync + ff foreground model, across three region-partition strategies: traditional splitting (left), FF-informed splitting (centre), and mixed splitting (right).
\textit{Top row:} fitted versus true region-averaged foreground sky temperatures, coloured by region index, with the dashed one-to-one line; pixel foreground RMSE values are $222$\,K, $219$\,K, and $148$\,K respectively.
\textit{Middle row:} reconstructed 21cm signal posterior mean (purple) with $1\sigma$, $2\sigma$, and $3\sigma$ credible bands, compared with the injected true signal (black); signal RMSE values are $0.0559$\,K, $0.0450$\,K, and $0.0131$\,K.
\textit{Bottom row:} normalised per-region parameter residuals $(\hat{\theta} - \theta_{\rm true})/\sigma$ for $A_{\rm sync}$ (red), $\beta_{\rm sync}$ (blue), and $A_{\rm ff}$ (green).
The mixed split achieves the lowest foreground RMSE while maintaining competitive signal recovery, demonstrating the benefit of component-specific region definitions.}
\label{fig:split_comparison}
\end{figure*}

Figure~\ref{fig:split_comparison} compares recovery performance for the three region-splitting strategies, when the sync + ff model is applied to the mixed-split dataset.
The top row shows fitted versus true region-averaged sky temperatures: the traditional and FF-informed splits achieve pixel RMSE values of $222$\,K and $219$\,K respectively, while the mixed split gives a markedly improved RMSE of $148$\,K.
The middle row shows the reconstructed 21cm signal posterior with $1\sigma$, $2\sigma$, and $3\sigma$ credible bands against the injected signal; signal RMSE values are $0.0559$\,K (traditional), $0.0450$\,K (FF-informed), and $0.0131$\,K (mixed).
The bottom row shows normalised per-region parameter residuals $(\hat{\theta} - \theta_{\rm true})/\sigma$ for $A_{\rm sync}$ (red), $\beta_{\rm sync}$ (blue), and $A_{\rm ff}$ (green).

The mixed split achieves the most accurate signal recovery and foreground mapping of all three methods.
By aligning the free-free region boundaries with the physical distribution of free-free emission, it eliminates the over-fitting of free-free amplitude in the innermost traditional region, which had been artificially inflated by the presence of higher-latitude, lower-emission pixels within the same region.
These results motivate using component-specific region definitions in the sync + ff model when physical knowledge of the emission distribution is available.

Mixed splitting was also applied to the \texttt{PySM3} data.
It produced no significant improvement in 21cm signal recovery, but reduced the overall foreground temperature RMSE from $359$\,K to $237$\,K compared with traditional splitting, indicating that mixed splitting improves foreground characterisation even when the signal itself cannot be recovered.

\section{Implications for Global 21\MakeLowercase{cm} Experiments}
\label{sec:discussion}

The optimal foreground model depends on the richness of the foreground spectral structure at REACH frequencies, which is not fully known.
The evidence of a fit using each model on real REACH observations will help to determine this; this is left for future work.
If the foregrounds are spectrally rich, then more complex models such as the curvature or sync + ff model are necessary to fit the 21cm signal, but this introduces additional degeneracies which reduce the accuracy of foreground parameter recovery, and therefore of the recovered foreground maps.
(The mixed region splitting described in Section~\ref{sec:splitting} can partially mitigate this).
Importantly, the default power law model fails to accurately fit data that deviates even slightly from a pure power law.

For foreground mapping, the sync + ff model may be preferred despite higher computational cost, as it recovers physical parameters that enable separate synchrotron and free-free maps, with value beyond 21cm cosmology.
The GPU-accelerated REACH pipeline presented in \cite{tutt2026optimising} makes this more computationally feasible for future work.
Mapping tolerates model-data mismatch better than signal extraction, as it requires only that bright, spectrally smooth emission be well characterised, rather than the mK-level residuals demanded by signal extraction.
However, the two objectives are not independent: regions where the foreground model is poorly constrained will degrade both the recovered map and the 21cm signal.

The curvature model provides a middle ground when spectral curvature is expected but full physical decomposition is not required. The curvature parameter $c$ encodes spectral structure that a pure power law cannot capture, despite the $\beta$/$c$ degeneracy that limits precise recovery of either parameter individually; for this reason, the model performs poorly on mismatched (i.e.\ not curved) data.

The variable amplitude model suits cases where regional brightness variations are expected (due to underlying map calibration errors) but spectral complexity is not, thus providing a middle ground between the power law and more complex models.

Changes in instrumental characteristics (e.g.\ by changing the antenna from the log-spiral considered here), or changes in the considered frequency range (which is dependent on calibration and our knowledge of the antenna beam), may shift the balance between models.

The sync + ff model recovers foreground parameters consistent with the physically motivated input simulations, within the uncertainties of each fit: synchrotron emission dominates across the REACH band with typical spectral indices $\beta_{\rm sync} \approx 2.5$ to $3.0$. This is consistent with expectations from Galactic synchrotron models \citep{Haslam1982,Reich1982,Zheng2017,Davies1998}.
Free-free emission becomes increasingly important toward higher frequencies and in regions near the Galactic plane.

We find significant spatial variations in both synchrotron and free-free parameters (Section~\ref{sec:component_maps}).
This reflects the complex structure of Galactic emission, including variations in cosmic ray electron spectra, magnetic field strength, and ionised gas distribution \citep{Davies1998,adam2016planck} - and as  $N_{\rm reg}$ is increased (beyond what was computationally feasible for all the combinations considered in this work), more of this structure is captured.

A notable result is the failure of all four models on \texttt{PySM3} data.
This is likely due to the fine spatial structure and spectral richness of the \texttt{PySM3 s1} model, as discussed in Section~\ref{sec:signal_extraction}, in combination with a deliberately simple model family (to minimise the effects of foreground parameter degeneracy).
This points to limitations of real REACH observations: if the true Galactic foreground at 50-170\,MHz exhibits spectral structure of comparable richness, none of the models tested here will be sufficient for 21cm signal recovery with $N_{\rm reg}=10$.

The difficulty of separating synchrotron and free-free emission is not unique to our pipeline.
The two components have similar spectral shapes at low frequencies and are strongly degenerate, and other studies have dealt with this in different ways.
\cite{adam2016planck} found that component separation required imposing a spatially constant synchrotron spectral index, rather than fitting it freely, to reduce the degeneracy, at the cost of mapping spatial variation in synchrotron spectral index.
For the same reason, \cite{peel2012template} use an H$\alpha$ template as a proxy for free-free emission, to avoid fitting it entirely, at the expense of increased uncertainty due to dust absorption.

For the analysis of real data from a global 21cm experiment, an initial fit with the power law model will give an indication of the spectral complexity of the real foregrounds, and therefore what model complexity is necessary. 
Follow-up analysis using the sync + ff model will allow the extraction of physical foreground parameters.
While more expensive, this provides valuable information about the foreground that complements the cosmological analysis.
A comparison of the Bayesian evidence for all models will assess which best describes the real data.

The mixed-splitting strategy introduced in Section~\ref{sec:splitting} demonstrates that component-specific region definitions reduce free-free over-fitting; a natural further extension is to use different foreground models in different regions, for instance, using a more complex model in the Galactic plane where spectral structure is richer.

Higher $N_{\rm reg}$ values would better capture spatial structure, and evidence-weighted averaging over a wider range of $N_{\rm reg}$ runs would further suppress region-boundary discontinuities in the recovered maps; both are feasible on real data where a single production run replaces the grid of tests carried out here.

Adding spectral curvature to the sync\,+\,ff model would increase physical interpretability, but preliminary tests showed strong degeneracies between $c$, $A_{\rm sync}$, and $A_{\rm ff}$; a higher number of live points would help, but this was also not practical across the full model grid in this work.

\section{Conclusions}
\label{sec:conclusions}
We have compared four foreground models - power law, variable amplitude power law, curvature, and sync + ff - against simulated REACH observations, assessing 21cm signal recovery, foreground parameter recovery, and component-separated sky mapping.

Model complexity must be matched to the spectral richness of the foreground. The power law model offers the cleanest 21cm signal extraction when foregrounds are spectrally simple: its low dimensionality avoids absorbing signal features and keeps computation tractable. Richer foreground structure requires more complex models, but additional parameters introduce degeneracies (particularly between spectral index and curvature, and between synchrotron and free-free amplitudes) that limit accurate parameter recovery.
Breaking these degeneracies in real REACH observations will likely require informative priors drawn from complementary multi-frequency data.

Out of these foreground models, the sync + ff model is a special case: it recovers physically meaningful synchrotron and free-free parameters that can be used to construct component-separated sky maps, with applications beyond 21cm cosmology \citep{padovani2021spectral}.
Synchrotron emission is well recovered across the sky, but free-free recovery is limited by the low signal-to-noise ratio of subdominant free-free emission between 50-170\,MHz, and by geometric mismatch between the fitted region boundaries and the physical concentration of free-free emission near the Galactic plane.
Therefore, physically motivated component models improve interpretability, but require tailored region-splitting to remain identifiable: grouping pixels by total sky brightness conflates regions of distinct physical origin, whereas partitioning each component by its own spatial distribution reduces foreground residuals without increasing the number of fitted parameters.
Evidence-weighted averaging over multiple $N_{\rm reg}$ runs further suppresses region-boundary discontinuities in the recovered maps.

A notable caveat is that all four models fail on \texttt{PySM3} data, whose spectral complexity exceeds what any of these models can capture with just 10 regions.
If the real sky at REACH frequencies exhibits comparable richness, a larger value of $N_{\rm reg}$, or external spectral constraints will be necessary before 21cm signal recovery is possible.

Foreground mapping and 21cm signal detection have different success criteria: mapping requires only that bright, spectrally smooth emission is well characterised, while signal detection demands mK-level residuals with no absorption of cosmological features \citep{chapman2019cosmic}.
Meaningful corrections to existing foreground maps can therefore be produced even before signal detection is achieved, and the continuous-frequency sky maps spanning 50-170~MHz that will emerge from REACH foreground fits are among the lowest-frequency all-sky maps ever produced.
At these frequencies, Galactic synchrotron and free-free emission are poorly constrained by existing surveys \citep{McKay2026}, and the spectral behaviour of diffuse emission between the frequencies of existing surveys remains largely uncharacterised.
REACH and other global 21cm experiments therefore occupy a dual role: the same observations that target the cosmological 21cm signal also probe Galactic physics that is inaccessible at higher frequencies.

\section*{Acknowledgements}

DR acknowledges the Science and Technology Facilities Council (STFC) Centre for Doctoral Training (CDT) in Data Intensive Science at the University of Cambridge for a PhD studentship.
EdLA acknowledges the support of UKRI STFC via grant ST/V004425/1 (Ernest Rutherford Fellowship).
DA and the wider REACH collaboration acknowledges the support of its funders: UKRI (EP/Y02916X/1 - ERC COG Horizon Europe Guarantee), The Kavli Institute for Cosmology in Cambridge, the South African National Research Foundation, Stellenbosch University, and the ALBORADA Trust Fund.
HTJB acknowledges support from the Kavli Institute for Cosmology Cambridge and the Kavli Foundation.
This work was performed using resources provided by the Cambridge Service for Data Driven Discovery (CSD3) operated by the University of Cambridge Research Computing Service.

\section*{Data Availability}

The data that support the findings of this study are available from the first author upon reasonable request.



\bibliographystyle{mnras}
\bibliography{references} 




\appendix

\section{Prior Specifications}
\label{app:priors}

The use of physically motivated priors, using existing foreground knowledge from low-frequency radio surveys is important at all stages of the analysis.
This is especially true for the sync + ff model, where weak free-free emission requires informative priors to constrain $A_{\rm ff}$.

Here, we provide prior specifications for all models used in this work.
All uniform priors are denoted $\mathcal{U}(a,b)$ for bounds $[a,b]$.

The prior ranges were chosen as follows.
The prior on $\beta_{\rm sync}$ is motivated by the spectral index distribution of the GSM \citep{deOliveiraCosta2008}, which yields values in the range $[2.5, 3.0]$ across the sky.
The prior on $\beta_{\rm ff}$ is motivated by the expected spectral index of optically thin free-free emission derived from Gaunt-factor free-free maps at REACH frequencies, which is consistent with $\beta_{\rm ff} \approx 2.1$ \citep{Condon2016,Xu2013,Dickinson2003}.
$A_{\rm sync}$ is centred close to unity by construction, and was refined over multiple trial fits to ensure the posterior was not truncated at either boundary.
$A_{\rm ff}$ was chosen such that the (subdominant) free-free contribution could reach up to half that of synchrotron emission in any given region.

\subsection*{Power law model}

For each region $i = 1, \ldots, N_{\rm reg}$:
\begin{align}
\beta_i &\sim \mathcal{U}(2.5,\, 3.0)
\end{align}

\subsection*{Variable amplitude power law model}

For each region $i$:
\begin{align}
A_i        &\sim \mathcal{U}(0.5,\, 1.5) \\
\beta_i    &\sim \mathcal{U}(2.5,\, 3.0)
\end{align}

The dimensionless amplitude $A_i$ scales the foreground contribution $(T_{230} - T_{\rm CMB})$ in each region, with the prior centred around unity to allow modest deviations from the spatial template.

\subsection*{Curvature model}

For each region $i$:
\begin{align}
A_i     &\sim \mathcal{U}(0.5,\, 1.5) \\
\beta_i &\sim \mathcal{U}(2.5,\, 3.0) \\
c_i     &\sim \mathcal{U}(-0.3,\, 0.3)
\end{align}

\subsection*{Sync + ff model}

For each region $i$:
\begin{align}
A_{{\rm sync}\mathit{i}}  &\sim \mathcal{U}(0.5,\, 1.5) \\
\beta_{{\rm sync}\mathit{i}} &\sim \mathcal{U}(2.5,\, 3.0) \\
A_{{\rm ff}\mathit{i}}    &\sim \mathcal{U}(-0.5,\, 0.5)
\end{align}

A variant of the sync + ff model in which $\beta_{\rm ff}$ is a free parameter was also considered, with
\begin{align}
\beta_{{\rm ff}\mathit{i}}   &\sim \mathcal{U}(2.0,\, 2.2).
\end{align}
This increases the complexity of the model to $4N_{\rm reg}$ free parameters per run, which increases the effect of degeneracy, leading to worse fits.
This motivated the decision to fix $\beta_{{\rm ff}\mathit{i}} = 2.1$, consistent with optically thin thermal bremsstrahlung \citep{Xu2013,Condon2016}.

\subsection*{21cm signal priors}

The fitted signal model is
\begin{equation}
T_{21}(\nu)=-A\exp\!\left[-\frac{(\nu-f_0)^2}{2\sigma^2}\right].
\end{equation}
We use:
\begin{align}
f_0   &\sim \mathcal{U}(50,\,200)\ {\rm MHz},\\
\sigma&\sim \mathcal{U}(10,\,20)\ {\rm MHz},\\
A     &\sim \mathcal{U}(0,\,0.4)\ {\rm K}.
\end{align}

\subsection*{Likelihood noise prior}

In the Gaussian likelihood of equation \ref{eq:likelihood}, $\theta_\sigma$ represents the (sampled) noise standard deviation (in K) across frequency channels in the fit.

We infer $\theta_\sigma$ jointly with foreground and signal parameters using a log-uniform prior:
\begin{equation}
\theta_\sigma \sim \mathcal{LU}(10^{-4},10^{-1})\ \mathrm{K}.
\end{equation}


\bsp	
\label{lastpage}
\end{document}